\documentclass[12pt,english,floatfix,nofootinbib,superscriptaddress,aps,prd,preprint]{revtex4-1}
\usepackage[utf8]{inputenc}
\usepackage{float}
\usepackage{array}
\usepackage{lipsum}
\usepackage{graphicx}
\usepackage{amsmath,amsthm,amsfonts,amssymb}
\usepackage{graphicx}
\usepackage[english]{babel}
\usepackage{color}
\usepackage{tensor}
\usepackage{esint}
\usepackage[dvips]{epsfig}
\usepackage[dvips]{graphicx}
\usepackage{float}
\usepackage{units}
\usepackage{textcomp}
\usepackage{mathrsfs}
\usepackage{amsmath}
\usepackage[makeroom]{cancel}
\usepackage{amssymb}
\usepackage{amsbsy}
\usepackage{amsfonts}
\usepackage{amssymb,mathrsfs,xcolor}
\usepackage{esint}
\usepackage{array}
\usepackage{graphicx}

\usepackage{hyperref}
\hypersetup{
    colorlinks,
    citecolor=blue,
    filecolor=green,
    linkcolor=purple,
    urlcolor=red,
}

\usepackage{slashed}

\newcommand{\ie}{\begin{equation}}
\newcommand{\fe}{\end{equation}}
\newcommand{\se}{\begin{eqnarray}}
\newcommand{\ff}{\end{eqnarray}}

\usepackage{hyperref}
\hypersetup{colorlinks,breaklinks,
			citecolor=[rgb]{0,0.0,1.0},
            urlcolor=[rgb]{0.0,0.0,1.0},
            linkcolor=[rgb]{0,0.5,0.9}}

\begin{document}

\title{Thermodynamics of massless particles in curved spacetime}

\author{A. A. Ara\'{u}jo Filho}
\email{dilto@fisica.ufc.br}

\affiliation{Universidade Federal do Cear\'a (UFC), Departamento de F\'isica,\\ Campus do Pici,
Fortaleza -- CE, C.P. 6030, 60455-760 -- Brazil.}

\affiliation{Departamento de Física Teórica and IFIC, Centro Mixto Universidad de Valencia--CSIC. Universidad de Valencia, Burjassot-46100, Valencia, Spain}




\date{\today}

\begin{abstract}

This work is devoted to study the behavior of massless particles within the context of curved spacetime. In essence, we investigate the consequences of the scale factor $C(\eta)$ of the Friedmann--Robertson--Walker metric in the Einstein--aether formalism to study photon--like particles. To do so, we consider the system within the canonical ensemble formalism in order to derive the following thermodynamic state quantities: spectral radiance, Helmholtz free energy, pressure, entropy, mean energy and the heat capacity. Moreover, the correction to the \textit{Stefan--Boltzmann} law and the equation of states are also provide. Particularly, we separate our study within
three distinct cases, i.e., $s=0,p=0$; $s=1,p=1$; $s=2,p=1$. In the first one, the results are derived \textit{numerically}. Nevertheless, for the rest of the cases, all the calculations are accomplished \textit{analytically} showing explicitly the dependence of the scale factor $C(\eta)$ and the Riemann zeta function $\xi(s)$. Furthermore, our analyses are accomplished in general taking into account three different regimes of temperature of the universe, i.e., the inflationary era ($T=10^{13}$ GeV), the electroweak epoch ($T=10^{3}$ GeV) and the cosmic microwave background ($T=10^{-13}$ GeV).

\end{abstract}

\maketitle


\section{Introduction}

Although the well--known theory of general relativity is certainly one of the most successful physical theories within theoretical and experimental viewpoints, there exist some alternative approaches to study gravity where the well--established Lorentz symmetry is no longer maintained  \cite{clifton2012modified,petrov2020introduction,olmo2007limit,jimenez2018born,olmo2011palatini}. Over the last years, many of them have received much attention, namely, Chern--Simons \cite{izaurieta2009standard,smith2008effects,aviles2018non,delsate2015initial,grumiller2008black,gomes2007induction,brito2008ambiguity,porfirio2016chern}, H\u orava--Lifshitz \cite{brandenberger2009matter,sotiriou2011hovrava,saridakis2010hovrava,kiritsis2010hovrava,li2009trouble,mukohyama2010hovrava,fonseca2013godel,furtado2011horava}, Einstein--aether \cite{jacobson2008einstein,jacobson2010extended,jacobson2004einstein,barausse2011black,lemoine2001stress,mazzitelli2011quantum,nacir2005renormalized,paliathanasis2020einstein}, Nieh--Yan \cite{nieh1982identity}, Bumblebee \cite{,maluf2021black,ovgun2019exact,casana2018exact,jha2021bumblebee,jha2021bumblebee1,oliveira2019quasinormal,jesus2020godel,delhom2021metric,gullu2021schwarzschild}, and non--local \cite{jaccard2013nonlocal,biswas2013nonlocal,modesto2013non,bergman2002nonlocal,calcagni2015nonlocal,buoninfante2019nonlocal} theories of gravity.

Particularly, the Einstein--aether theory has its Lorentz symmetry broken due to the insertion of a timelike vector field (a unitary one)-- known as ``aether"-- in the widely--spread Einstein--Hilbert action \cite{jacobson2001gravity,carruthers2011cosmic,zlosnik2007modifying,ghosh2007random,carroll2004lorentz}. The quadratic terms in the kinetic part of its respective lagrangian clearly changes the usual gravitational action ascribed to the general relativity. Being a general second--order theory governed by the metric $g_{\mu\nu}$ and the vector field $u_{\mu}$ (the aether field), the Einstein--aether action does not possess more than two derivatives \cite{ghosh2007random,carroll2004lorentz}.

One of the most notable features concerning this theory is that it may be viewed as a particular limit of the H\u orava–Lifshitz theory of gravity \cite{paliathanasis2020einstein}. In other words, this imply that the solutions of Einstein--aether theory are compatible with H\u orava–Lifshitz approach. Nevertheless, the reciprocal way is not ensured-- generically, it is not guaranteed that the solutions of H\u orava--Lifshitz theory be compatible with Einstein--aether formalism. Moreover, despite of having an agreement between solutions of these theories (Einstein--aether into H\u orava--Lifshitz), a complete correspondence when the whole field equations of motion are taken into account is not assured \cite{sotiriou2011hovrava}.

On the other hand, the black hole solutions involving Einstein--aether theory of gravity are discussed, which was shown that the outside solution is similar to the one encountered in general relativity, the Schwarzschild solution \cite{eling2006black}. In contrast, the solutions which lie inside disagree with those exhibited in general relativity. More so, with fluid sources characterizing neutron stars, it was accomplished an investigation in the context of spherically symmetric spacetimes \cite{eling2007neutron}. A concise examination of Einstein--aether theory involving spherically symmetric spacetimes with perfect fluids was also provided \cite{coley2015spherically}.

Essentially, regarding diverse perfect fluid models in Ref. \cite{coley2015spherically}, it is investigated the local stability concerning the gravitational field equations taking into account inhomogeneous cosmological scenarios and compact objects. Furthermore, it is worth arguing that perfect fluid models within Kantowski--Sachs Einstein--aether theory were examined \cite{latta2016kantowski}. The scalar field was also considered within the context of static spherically symmetric solutions \cite{coley2019static,leon2020static} and, recently, for homogeneous and anisotropic spacetimes, the exact solutions were performed \cite{roumeliotis2019reduced,roumeliotis2020exact}. In parallel, the Einstein--aether theory has also been addressed as a dark--energy candidate in order to elucidate the well--known acceleration phase of the universe \cite{meng2012einstein,rani2019cosmological}. On the other hand, a robust exploration of an inflationary model in the context of Lorentz violation was provided by Kanno and Soda \cite{kanno2006lorentz}. In other words, it was considered a nonminimally coupling of a scalar field with the aether field \cite{donnelly2010coupling}. In this formalism, considering moreover Friedmann--Robertson--Walker (FRW) spacetime, many studies have been carried out \cite{barrow2012some,solomon2014inflationary,paliathanasis2019dynamics,paliathanasis2020analytic,sandin2013stability,alhulaimi2017spatially,van2018kantowski,mohandas2020kantowski,paliathanasis2020dynamics}.

One of the notable features of theories involving Lorentz symmetry breaking is undoubtedly their respective thermal aspects. Investigations in such direction can supply further knowledge concerning primordial stages of the expansion of the Universe, in which the size at these scenarios is in agreement with characteristic scales of Lorentz symmetry breaking \cite{amelino2001testable}. The procedure to study the thermodynamic properties in Lorentz violating theories has pioneer been proposed by Colladay and McDonald \cite{colladay2004statistical}. Subsequently, many studies have been accomplished  utilizing such procedure \cite{petrov1,petrov2,euso,aa2022particles,alfieres1,eucommaluf,araujo2022particles,gomes2010free,casana2008lorentz,casana2009finite,anacleto2018lorentz,das2009relativistic,aguirre2021lorentz,furtado2020effects,chacon2011statistical,thermal1,thermal2,altschul2007limits,castellanos2010stability}. However, in the context of Einstein--aether theory with Friedmann--Robertson--Walker spacetime there is a lack in the literature providing the investigation of its respective thermal aspects ascribed to this formalism. Mainly, the physical implications of the thermodynamic properties in the context of Einstein--aether theory might depict some new fingerprints of a new phenomenon that might be confronted with observatory data as long as it is available in order to obtain any trace of Lorentz violation.

In order to overcome this situation and provide a toy model for additional studies regarding cosmology, this work is aimed at exploring the behavior of massless modes in curved spacetime, i.e., we investigate the impact of the scale factor $C(\eta)$ of the Friedmann--Robertson--Walker metric in the Einstein--aether approach to study the thermal aspects of photon--like particles based on the procedure used by Amelino--Camelia \cite{amelino2001testable}. For accomplishing it, we regard the canonical ensemble formalism to proceed forward with our calculations concerning the thermodynamic functions. In order to provide a concise investigation, we separate our study in tree distinct cases: $s=0,p=0$; $s=1,p=1$; $s=2,p=1$. The first case, the results are carried out in a \textit{numerical} manner; nevertheless, for the latter two ones, all results are accomplished \textit{analytically} exhibiting an explicit dependence of the scale factor $C(\eta)$ and the Riemann zeta function $\xi(s)$. More so, three different regimes of temperature of the universe are also considered in our calculations, i.e., the inflationary era ($T=10^{13}$ GeV), the electroweak epoch ($T=10^{3}$ GeV) and the cosmic microwave background ($T=10^{-13}$ GeV).


\section{Modified dispersion relation in curved spacetime}

In curved spacetimes, there exist many ways to introduce modified dispersion ralations. As it is already well--known, a generic modified dispersion relation makes the invariance of local Lorentz frame no longer maintained. Nevertheless, the covariance can be preserved by the insertion of a dynamical vector field $u_{\mu}$, i.e., the aether field, in which has the constraint of taking a nonzero timelike value, $u_{\mu}u^{\mu} = - 1$. In this way, we shall present a general covariant form within the framework of the Einstein--aether theory \cite{jacobson2001generally,jacobson2001gravity,lemoine2001stress} in order to accomplish further analysis.

Regarding the well--known semiclassical approximation, both the respective spacetime metric $g_{\mu\nu}$ and the aether field are considered to be classical ones. Moreover, the associated action to the scalar field $\varphi$ can properly be written as follows:
\ie
\mathcal{S}_{\varphi} = - \frac{1}{2} \mathrm{d}^{n}x \sqrt{-g} \left\{ \partial^{\mu}\varphi \partial_{\mu}\varphi + 2 [\lambda (g^{\mu\nu}u_{\mu}u_{\nu} + 1) - l F^{\mu\nu}F_{\mu\nu}] + 2 \sum_{s,p\leqslant s} b_{sp} (\mathcal{D}^{2s}\varphi) (\mathcal{D}^{2p}\varphi),  \right\} \label{themainlagrangian}
\fe
where $b_{sp}$ and $l$ are arbitrary constants; $\lambda$ is the Lagrangian multiplier; the tensor $F^{\mu\nu}$ is defined as $F^{\mu\nu} \equiv \nabla_{\mu} u_{\nu} - \nabla_{\nu} u_{\mu} $; and $g_{\mu\nu}$ is the metric. Furthermore, the operators $\mathcal{D}^{2p}$ are defined as $\mathcal{D}^{2p} \equiv \mathcal{D}^{\mu  1}\mathcal{D}_{\mu 1}\mathcal{D}^{\mu 2}\mathcal{D}_{\mu 2} ... \mathcal{D}^{\mu p}\mathcal{D}_{\mu p}$. As one can expect, the action displayed in Eq. (\ref{themainlagrangian}) keeps the general principal of covariance, which will clearly guarantee the conservation of the stress--energy $T^{\mu\nu}$. It is worth mentioning that if the vector field $u_{\mu}$ is geodesic, it follows immediately that the tensor $F_{\mu\nu} = 0$. The Einstein--aether theory is a modification of general relativity that introduces a background field known as the aether. It is a preferred frame of reference that breaks the Lorentz symmetry spontaneously. Such a feature is required in order to preserve the Bianchi identity in gravitational scenarios for instance \cite{bluhm2008spontaneous}.

One of the motivations behind the Einstein--aether theory is to address some of the unresolved issues in general relativity, such as the dark matter and the cosmological constant problems. These problems arise from the observation that the observed behavior of gravity on large scales does not match the predictions of general relativity.

In addition, another motivation for the theory displayed in  Eq. (\ref{themainlagrangian})  is to provide a more complete picture of the fundamental laws of physics by incorporating quantum corrections. To achieve this, there exists Einstein--aether like term added by higher--order derivatives, which allow us to obtain finite quantum corrections, considering also particle creation process \cite{lemoine2001stress}.

Here, one important feature to this approach is considering a globally hyperbolic spacetime $\mathcal{M}$; thereby, we can properly foliated it in term of a three dimensional spacelike hyper surface $\Sigma$ as $\mathcal{M} = \Sigma \times \mathbb{R}$ \cite{hawking1973large,wald1994quantum,wald2010general,carroll2019spacetime,padmanabhan2010gravitation}. Now, let us define the projection operator $\perp_{\mu\nu}$ \cite{nacir2005renormalized,lemoine2001stress,mazzitelli2011quantum} in order to step toward our calculations. In this sense, the projection operator is given by
\ie
\perp_{\mu\nu} = g_{\mu\nu} + u_{\mu}u_{\nu}.
\fe
With this definition, we can write the invariant of the spacetime as 
\ie
\mathrm{d}s^{2} = g_{\mu\nu}\mathrm{d}x^{\mu}\mathrm{d}x^{\nu} = -(u_{\mu}\mathrm{d}x^{\mu})^{2} + \perp_{\mu\nu}\mathrm{d}x^{\mu}\mathrm{d}x^{\nu}.
\fe
Now, our next step is building up the stress--energy tensor in which can be an alternative to also acquire the modified dispersion relations related to the theory \cite{neves2021dispersion,diles2020third,gedalin1995generally}. To obtain such tensor, we have to vary the Eq. (\ref{themainlagrangian}) in terms of the metric tensor $g_{\mu\nu}$
\ie
\delta \mathcal{S}_{\varphi} = - \frac{1}{2} \int \mathrm{d}^{4}x \sqrt{-g} \, T^{\mu\nu} \delta g_{\mu\nu} = \frac{1}{2} \int \mathrm{d}^{4}x \sqrt{-g} \, T_{\mu\nu} \delta g^{\mu\nu}.
\fe
After some steps of calculation which can be encountered in Ref. \cite{lemoine2001stress}, we have the form of the stress--energy tensor:
\ie
\begin{split}
T_{\mu\nu} = \, &\partial_{\mu}\varphi \partial_{\nu}\varphi -\frac{1}{2} (\partial_{\alpha} \varphi \partial^{\nu} \varphi)g_{\mu\nu} - \sum_{s,p} b_{sp} \mathcal{D}^{2s}\varphi \mathcal{D}^{2p} \varphi g_{\mu\nu} + 2 \lambda u_{\mu}u_{\nu} \\ 
& + 2 \left[ u_{(\mu} \nabla^{\alpha}u_{\nu)}\nabla_{\alpha} \varphi - u_{(\mu} \nabla_{\nu)}u^{\alpha}\nabla_{\alpha}\varphi - a_{(\mu}\nabla_{\nu)} \varphi   \right. \\
& \left. \frac{1}{2} \perp_{\mu\nu} \Box \varphi + \frac{1}{2}g_{\mu\nu} (\ddot{\varphi} + \perp_{\mu\nu} \nabla_{\mu} u_{\nu} \dot{\varphi})
\right] \Upsilon - 2 \nabla_{(\mu}\Upsilon\, \nabla_{\nu)} \varphi \\
& -2 \dot{\Upsilon} u_{(\mu}\nabla_{\nu)} \varphi - 2 \dot{\varphi} u_{(\mu}\nabla_{\nu)}\Upsilon + \perp_{\mu\nu} \nabla_{\alpha}\Upsilon\, \nabla^{\alpha} \varphi + g_{\mu\nu} \dot{\Upsilon} \dot{\varphi}\\
& -4 l F_{\mu\lambda} F\indices{^\lambda_\nu} + l g_{\mu\nu} F^{\alpha\beta}F_{\alpha\beta} 
\label{stress-energy}
\end{split}
\fe
where $\Upsilon (\varphi) \equiv \sum b_{sp} (s+p) \mathcal{D}^{2s+2p-2} \varphi$. As a canonical procedure, for a comoving observer with $u_{\mu}$, we may also get the information about the pressure and energy density:
\ie
\tilde{p} \equiv \frac{1}{3} \perp^{\mu\nu}T_{\mu\nu}, \,\,\,\,\,\, \tilde{\rho} \equiv u^{\mu}u^{\nu} T_{\mu\nu}.
\fe
In this way, we obtain
\ie
\begin{split}
\tilde{p} = \frac{1}{3} &\left\{  \dot{\varphi}^{2} - \frac{1}{2} \nabla_{\alpha} \varphi \nabla^{\alpha} \varphi  - 3 \sum_{s,p} b_{sp} \mathcal{D}^{2s}\varphi \mathcal{D}^{2p}\varphi \,  - \perp_{\mu\nu} \left( -4 l F_{\mu\lambda} F\indices{^\lambda_\nu} + l g_{\mu\nu} F^{\alpha\beta}F_{\alpha\beta} \right)  \right. \\
& \left. + (3 \mathcal{D}^{2}\varphi + a^{\alpha} \nabla_{\alpha} \varphi   )\Upsilon(\varphi)  +  \nabla_{\alpha} \Upsilon \,\nabla^{\alpha} \varphi + \dot{\Upsilon}\dot{\varphi} \right\},
\end{split}
\fe
and, 
\ie
\tilde{\rho} = \dot{\varphi}^{2} + \frac{1}{2} \nabla_{\mu}\varphi\nabla^{\alpha}\varphi + \sum_{s,p} b_{sp} \mathcal{D}^{2s}\varphi \mathcal{D}^{2p}\varphi + u^{\mu}u^{\nu}\left( -4 l F_{\mu\lambda} F\indices{^\lambda_\nu} + l g_{\mu\nu} F^{\alpha\beta}F_{\alpha\beta} \right) + 4 l u^{\mu}\nabla^{\nu}F_{\nu\mu}.
\fe
We may particularize above results taking into account the Friedmann--Robertson--Walker spacetime. Being an exact solution of Einstein's field equations of general relativity, the well--known Friedmann--Robertson--Walker metric outlines an isotropic, homogeneous, and expanding universe. Within a strict mathematical viewpoint, it has to be
path--connected instead of being a fundamentally simply connected one \cite{munkres2000topology,munkres2018elements,kelley2017general,bredon2013topology,mendelson1990introduction,gamelin1999introduction,lee2010introduction}. It is worth mentioning that the general remarkable features of this metric come essentially from the geometric properties of its isotropy as well as its homogeneity. Moreover, the well--known Einstein's field equations are necessary to give us the scale factor $C(\eta)$ of the universe as a function of the proper time $\eta$. The metric is 
\ie
\mathrm{d}s^{2} = g_{\mu\nu} \mathrm{d}x^{\mu} \mathrm{d}x^{\nu} =  -\mathrm{d}\eta^{2} + C(\eta)^{2} \mathrm{d} \Psi^{2},
\fe
where $\Psi$ lies over a 3--dimensional space of uniform curvature. This metric is one of the most studied ones whenever one deals with cosmological scenarios \cite{pavon1991causal,stewart1990perturbations,boyanovsky1997scalar,szulc2007closed,lewis2000efficient,melia2016physical,melia2017zero,melia2019lapse,bikwa2012photon,cai2012emergence,jain2007analog,aref2008dynamics,molina1999minimal,zhu2009corrections,faraoni2003total}. Now, for the sake of obtaining the modified dispersion relation, we derive the field equation to $\varphi$. Let us consider $\chi \equiv C\varphi$; varying the action in Eq. (\ref{themainlagrangian}) with respect to $\chi$, we obtain
\ie
\chi'' - \frac{C''}{C}\chi - \Delta \chi = - 2 C^{2} \sum_{s,p} \frac{b_{sp}}{C^{2(s+p)}} \, \Delta^{2(s+p)} \chi,
\fe
where the prime indices represents the derivative with respect to the proper time $\eta$. Here, considering the plane wave solutions $\chi(\eta,{\bf{x}}) = \chi(\eta) e^{ik\cdot x}$, for a scalar field $\varphi$ in $n$--dimensional Robertson–Walker metrics, the Fourier modes of the scaled field $\chi=C^{(n-2)/4}(\eta)\,\varphi$ satisfy
\ie
\chi(\eta)'' + \chi \left[ \omega^{2} - \frac{C''}{C}    \right] = 0,
\fe
which leads to
\ie
\omega_{k}^{2} = {\bf{k}}^{2} + C(\eta) \left[  m^{2} + \sum_{s,p\leqslant s} (-1)^{s+p}\,b_{sp} \left(  \frac{{\bf{k}}}{\sqrt{C(\eta)}} \right)^{2(s+p)}          \right],
\label{generaldispersionrelation}
\fe
where it represents a Amelino--Camelia like dispersion relations. As it is straightforward to realize, when $s$ and $p$ run, Eq. (\ref{generaldispersionrelation}) turns out to describe a variety of the gravitational theories, e.g., Hořava--Lifshitz gravity. Additionally, the study of modified dispersion relations is motivated by the desire to understand the impact of such modifications on the properties of quantum fields, particularly in the context of cosmology and inflationary models. Previous studies have shown that certain modifications to the dispersion relation, such as the introduction of complex values, can lead to problematic situations in the context of quantum field theory \cite{2,3}. However, recent work has demonstrated that dispersion relations that break adiabaticity may lead to modifications of the power spectrum, but at the expense of the creation of a possibly large amount of energy density \cite{14,15,16}.

Furthermore, such a study has important implications for understanding the behavior of trans--Planckian modes in inflationary cosmology \cite{10,11,12,13}, as well as the possibility of explaining the form of vacuum energy seen in the Universe today through the use of gravitons with super--Planck momenta but frequencies much smaller than the Hubble expansion rate \cite{20}. In addition, it is important to mention that very recently in the literature, it is studied the effects of modified dispersion relations on free Fermi gas with application to astrophysics \cite{santos2021effects}.

In the next sections, we shall derive the thermodynamic behavior of our system considering three different configurations of Eq.(\ref{generaldispersionrelation}) based on the procedure proposed by \cite{amelino2001testable}. The present manuscript is devoted to study the remarks of the corresponding theory regarding exclusively this dispersion relation. Furthermore, we regard three distinct temperatures of the universe, i.e., the inflationary era ($T=10^{13}$ GeV), the electroweak epoch ($T=10^{3}$ GeV) and the cosmic microwave background ($T=10^{-13}$ GeV). Moreover, as we can verify from Eq. (\ref{generaldispersionrelation}), we may also obtain indefinitely dispersion relations. Nevertheless, our purpose in this work is particularizing them in some different cases in order to provide a concise analysis concerning the thermodynamic aspects of the theory encountered in Eq. (\ref{themainlagrangian}).


\section{Thermodynamic state quantities}

Modifications to the dispersion relation could potentially have an impact on the behavior of particles in the early universe, and therefore, on the \textit{Cosmic microwave background} (CMB). In particular, some studies have suggested that modified dispersion relations could lead to changes in the propagation of photons in the early universe, which could in turn affect the polarization patterns observed in the CMB \cite{b,bb,bbb,bbbb}. By studying the thermodynamic properties of such systems, we might be able to gain insights into the underlying physical processes that generated the CMB and how they are affected by modifications to the dispersion relation.

More so, it is worth noting that this is a complex and active area of research, and the relationship between modified dispersion relations, thermodynamics, and the CMB is not yet fully understood. Nevertheless, in Ref. \cite{JCAP}, the authors addressed how the modified dispersion relation of graviton affected the \textit{cosmic microwave background} power spectra, in particular the $B$--mode polarization. In our case, further analysis must be accomplish in order to properly ensure this feature.

Initially, we construct of the well--known partition function for the sake of obtaining the following thermodynamic properties: Helmholtz free energy, mean energy, pressure, entropy, and heat capacity. The spectral radiance(the black body radiation), the equation of states and the correction to the \textit{Stefan--Boltzmann} law are also provided via such formalism. In this sense, we use a traditional method for doing so; we use the concept of the number of accessible states of the system \cite{greiner2012thermodynamics,reif2009fundamentals,pathria2011statistical}. Generically, it can be written as
\ie
\Omega(E) = \frac{\Gamma}{\pi^{2}} \int^{\infty}_{0} \mathrm{d} {\bf{k}} |{\bf{k}}|^{2}, \label{ms2}
\fe
where $\Gamma$ is considered the volume of the thermal bath. More so, all of our calculations provided in the next sections, i.e., involving all the thermodynamic state quantities such as mean energy, entropy, Helmholtz free energy, entropy, pressure, spectral radiance, and the heat capacity, will be accomplished in a ``per volume" approach.

In order to provide a better comprehension for the reader, we provide a general definition of the partition function for an indistinguishable spinless gas \cite{greiner2012thermodynamics}: 
\ie
Z(T,\Gamma,N) = \frac{1}{N!h^{3N}} \int \mathrm{d}q^{3N}\mathrm{d}p^{3N} e^{-\beta H(q,p)}  \equiv \int \mathrm{d}E \,\Omega(E) e^{-\beta E}, \label{partti1}
\fe
where $q$ is the generalized coordinates, $p$ is the generalized momenta, $\beta = 1/\kappa_{B} T$, $N$ is the number of particles, $H$ is the Hamiltonian of the system, $h$ is the Planck's constant. Notice that the main issue here is obtaining the accessible states of the system $\Omega(E)$; with it, the obtainment of the partition function turns out to be a straightforward task \cite{oliveira2019thermodynamic,oliveira2020relativistic,oliveira2020thermodynamic,reis2020does,reis2021fermions,pacheco2014three,reis1,reis2}. 
However, Eq. (\ref{partti1}) does not take into account the spin of the particles that we are working on. In our case, we have to account for the massless particles(bosons) as follows: 
\ie
\mathrm{ln}[Z] = \int \mathrm{d}E \,\Omega(E) \mathrm{ln} [ 1- e^{-\beta E}],
\fe
where the factor $\mathrm{ln} [ 1- e^{-\beta E}]$ accounts for the Bose--Einstein statistics. With it, the following thermal quantities can properly be derived
\ie
\begin{split}
 & F(\beta,C(\eta))=-\frac{1}{\beta} \mathrm{ln}\left[Z(\beta,C(\eta))\right], \\
 & U(\beta,C(\eta))=-\frac{\partial}{\partial\beta} \mathrm{ln}\left[Z(\beta,C(\eta))\right], \\
 & S(\beta,C(\eta))=k_B\beta^2\frac{\partial}{\partial\beta}F(\beta,C(\eta)), \\
 & C_V(\beta,C(\eta))=-k_B\beta^2\frac{\partial}{\partial\beta}U(\beta,C(\eta)).
\label{thermodynamicproperties}
\end{split}
\fe 
All of them, the spectral radiance, the correction to the \textit{Stefan--Boltzmann} law as well as the equation of states will be addressed in the next sections. Hereafter, our purpose will be focused on studying the behavior of our system regarding particularly the cases coming from Eq. (\ref{generaldispersionrelation}), i.e., we provide a robust investigation of the thermal aspects concerning some particular cases. In essence, we consider three different configurations of the system within distinct ranges of temperature as follows.

\section{The simplest case: $s=0,  p=0$}

We start with the simplest case obtained from Eq. (\ref{generaldispersionrelation}) pointed out previously. In other words, we consider the case where we have the following configuration: $s=0, p=0$. With it, we obtain the dispersion relation: $E^{2}={\bf{k}}^{2} + C(\eta)$. However, for this configuration of the system, the accessible states are given by
\ie
\Omega(C(\eta)) = \frac{\Gamma}{\pi^{2}} \int^{\infty}_{0} [E^{2}-C(\eta)]\mathrm{d} E,
\fe
where $\Gamma$ is the volume of the thermal reservoir. In order to accomplish our calculations, we use the formalism of the canonical ensemble; in this way, we use the partition function:
\ie
\mathrm{ln} Z[\beta,C(\eta)] = - \frac{\Gamma}{\pi^{2}} \int^{\infty}_{0} [E^{2}-C(\eta)] \mathrm{ln}(1-e^{-\beta E}) \mathrm{d} E.
\fe
With this quantity, all the thermodynamic functions can be derived in the next sections, e.g., spectral radiance, mean energy, entropy and heat capacity. The equation of states as well as the correction to the \textit{Stefan--Boltzmann} law will also be provided. It worth mentioning that all thermal quantities will be studied within three distinct regimes, i.e., inflationary era ($T=10^{13}$ GeV), electroweak epoch ($T=10^{3}$ GeV), and cosmic microwave background ($T=10^{-13}$ GeV). As a straightforward approach, we first compute the spectral radiance.

\subsubsection{The spectral radiance}

\begin{figure}[tbh]
  \centering
  \includegraphics[width=8cm,height=5cm]{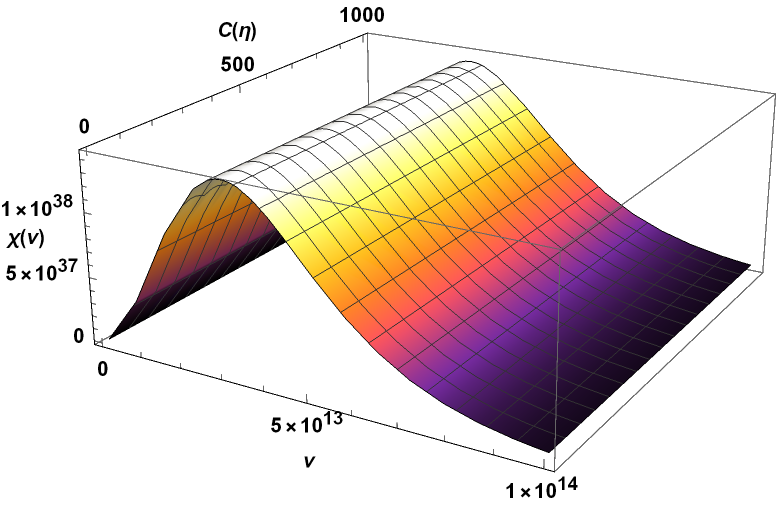}
  \includegraphics[width=8cm,height=5cm]{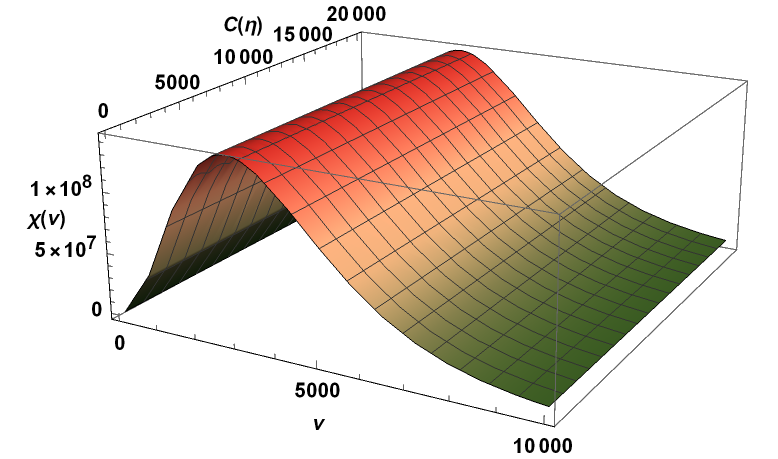}
  \includegraphics[width=8cm,height=5cm]{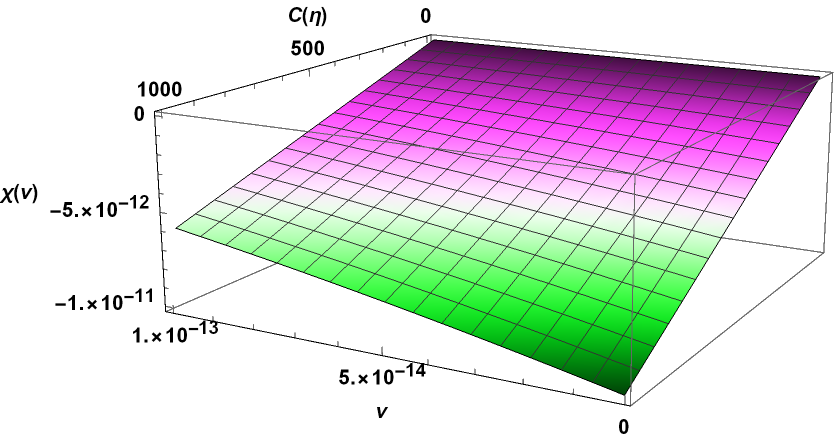}
  \caption{The spectral radiance for different configurations of temperature: top left (inflationary era), top right (electroweak epoch), and bottom (CMB)}\label{trivialcase-spectralradiance}
\end{figure}

As it is well--known, the emanation of heat and light coming from electromagnetic radiation changes with temperature. Such aspect is called black body radiation in the literature; the acquirement of a concise theory, which overcame a deadlock through low and high frequencies in 1900's, was proposed by Planck \cite{kragh2000max}. Nevertheless, in the context of a modified dispersion relation presented in Eq. (\ref{generaldispersionrelation}), such electromagnetic radiation might reveal some fingerprints of a particular theory which breaks the Lorentz symmetry. In this direction, we calculate it as follows
\ie
\chi(\beta,C(\eta),\nu) = \frac{h \nu  \left[h\nu ^2-C(\eta)\right]  e^{-\beta  h\nu} }{\pi^{2} [1-e^{-\beta  h\nu}]},
\fe
in which we have considered that $E=h\nu$, where $h$ is the Planck constant, and $\nu$ is the frequency. For a better comprehension, we plot the spectral radiance in Fig. \ref{trivialcase-spectralradiance} in order to analyse the behavior of this physical quantity. Particularly, we focus on investigating the consequences of different temperatures in this scenario, e.g., $T=10^{13}$ GeV (inflationary epoch), $T=10^{3}$ GeV (electroweak era), and $T=10^{-13}$ GeV (cosmic microwave background). We see that its shape is sensible to the scale factor $C(\eta)$ with a critical value around $C(\eta)=2\text{x}10^{5}$; after this value, the system seems to show instability as one can verify from the same plot.

Another remark worth exploring is what are the consequences of scale factor $C(\eta)$ in our thermodynamic system specifically concerning the correction to the \textit{Stefan--Boltzmann} law. To accomplish such analysis, we address the next section.

\subsubsection{The Stefan--Boltzmann law}

\begin{figure}[tbh]
  \centering
  \includegraphics[width=8cm,height=5cm]{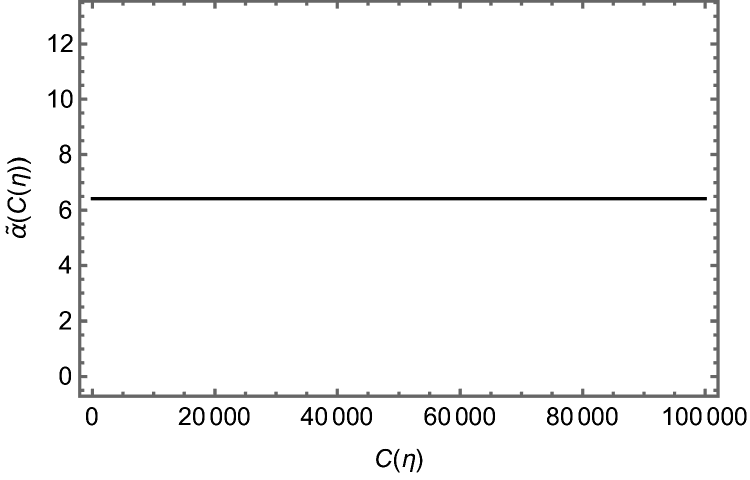}
  \includegraphics[width=8cm,height=5cm]{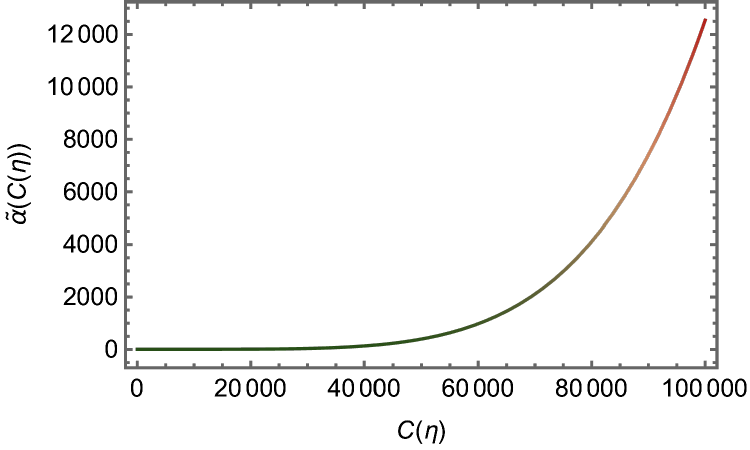}
  \includegraphics[width=8cm,height=5cm]{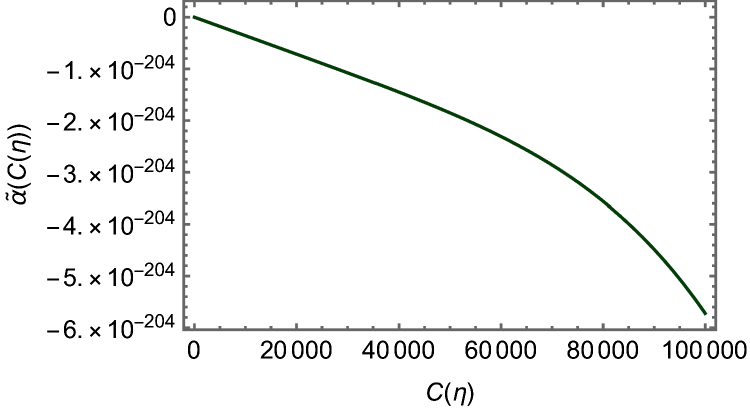}
  \caption{The correction to the \textit{Stefan--Boltzmann} law considering three different ranges of temperature: top left (inflation era), top right (electroweak epoch), and bottom (CMB)}\label{sblaw}
\end{figure}

This section is devoted to investigate how the scale factor $C(\eta)$ interferes in the \textit{Stefan--Boltzmann} law. For the sake of providing a proper answer to this point, we consider
\ie
\tilde{\alpha}(\beta,C(\eta)) \equiv U(\beta,C(\eta)) \beta^{4}.
\fe
A complete analysis to the above parameter $\tilde{\alpha}(\beta,C(\eta))$ that accounts for the correction to the \textit{Stefan--Boltzmann} law regarding three different temperature regimes is shown in Fig. \ref{sblaw}. As one can naturally expect, if $C(\eta) \rightarrow 0$, we will recover the well-established result in the literature, i.e., $\alpha = \pi^{2}/15$ \cite{huang2009introduction,salinas2001introduction}.

We see that when the high temperature regime ($T=10^{13}$ GeV) is taken into account, the parameter $\tilde{\alpha}(\beta,C(\eta))$ keeps the same. Also, for the temperature in the electroweak epoch ($T=10^{3}$ GeV), the plot resembles the well-known results initially proposed in the literature, i.e., $u_{S} \sim T^{4}$. On the other hand, when the temperature approaches to cosmic microwave background ($T= 10^{-13}$ GeV), the system seems to indicate instability.

\subsubsection{The Helmholtz free energy}

\begin{figure}[tbh]
  \centering
  \includegraphics[width=8cm,height=5cm]{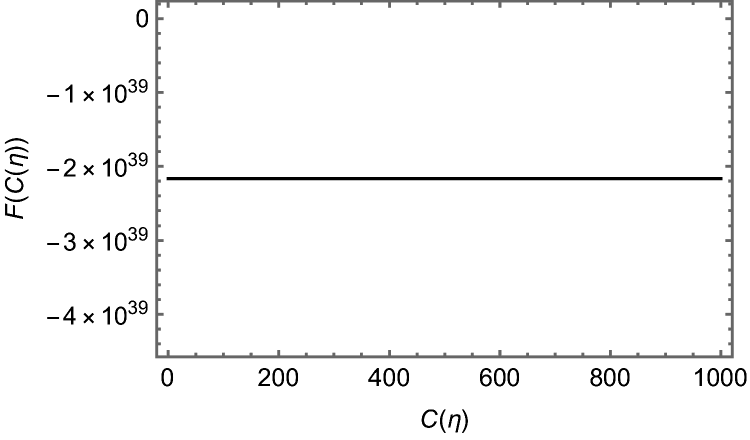}
  \includegraphics[width=8cm,height=5cm]{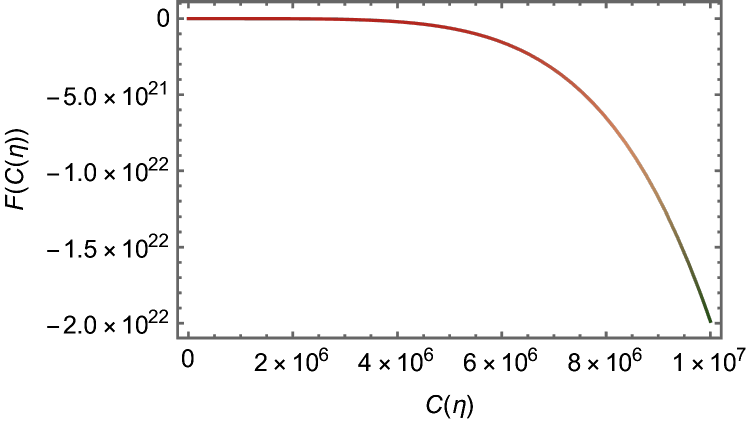}
  \includegraphics[width=8cm,height=5cm]{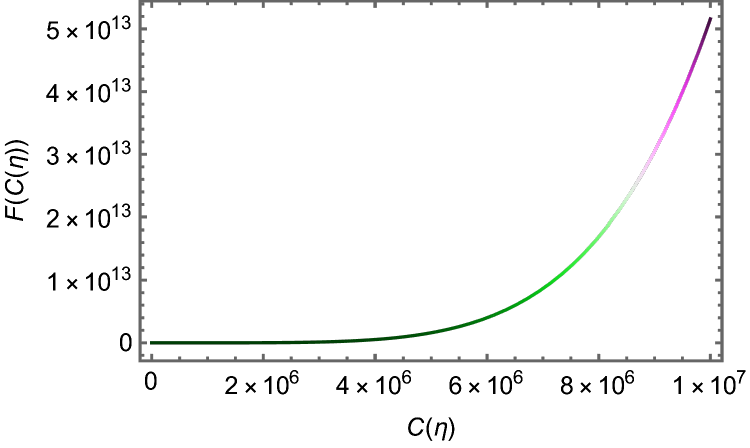}
  \caption{The Helmholtz free energy for distinct temperatures: top left (inflation era), top right (electroweak epoch), and bottom (CMB)}\label{Helmholtz}
\end{figure}

For the simplest case, the Helmholtz free energy can be written as 
\ie
F(\beta,C(\eta)) = \frac{1}{\beta \pi^{2}} \int^{\infty}_{0} [E^{2}-C(\eta)] \mathrm{ln}(1-e^{-\beta E}) \mathrm{d}E.
\fe
The corresponding behavior of the Helmholtz free energy is displayed in Fig. \ref{Helmholtz}; for high temperature scale, our system tends to maintain the same value of $F(C(\eta))$ when the scale factor $C(\eta)$ increases. The parameter $F(C(\eta))$ decreases when the scale factor $C(\eta)$ increases. Nevertheless, for very low temperatures, the Helmholtz free energy increases for different values of $C(\eta)$, which such a behavior might indicate instability for this regime of temperature.

\subsubsection{The equation of states}

\begin{figure}[tbh]
  \centering
  \includegraphics[width=8cm,height=5cm]{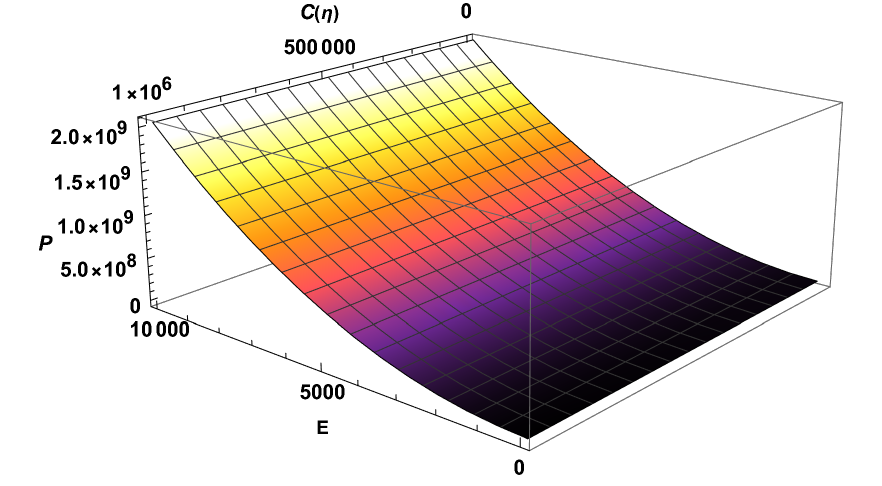}
  \includegraphics[width=8cm,height=5cm]{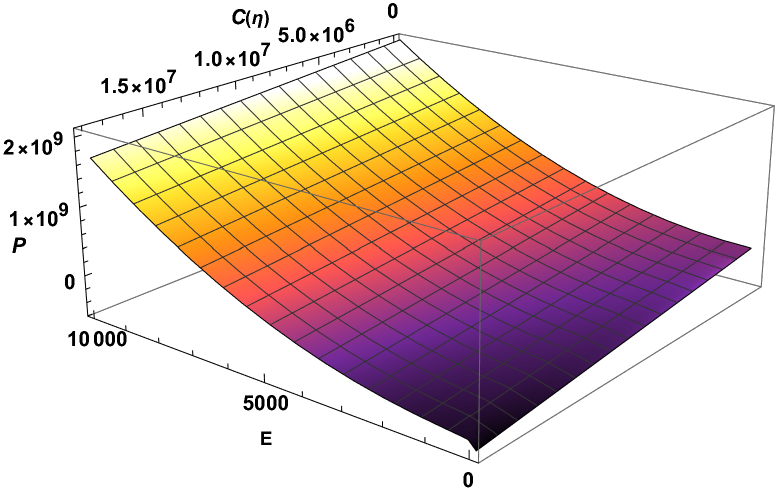}
  \includegraphics[width=8cm,height=5cm]{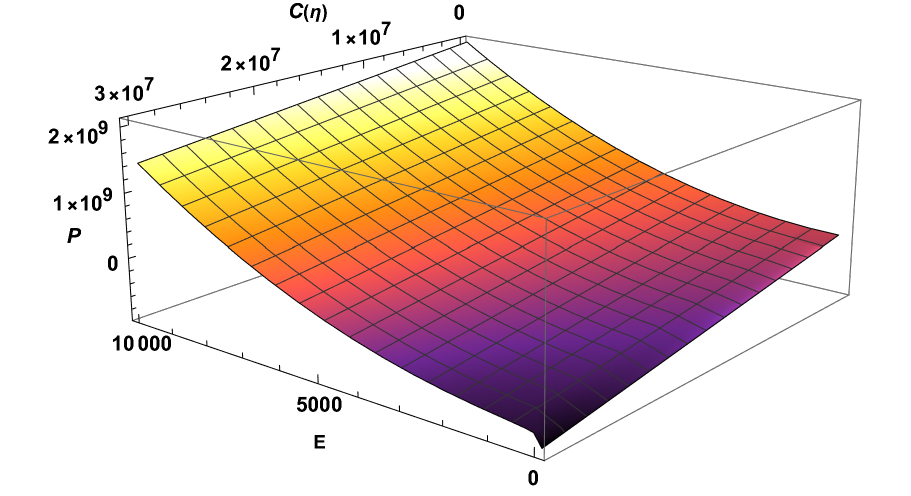}
  \caption{The modifications of the equation of states due to the scale factor $C(\eta)$ considering the inflationary regime of temperature }\label{equationofstatesinflation}
\end{figure}

\begin{figure}[tbh]
  \centering
  \includegraphics[width=8cm,height=5cm]{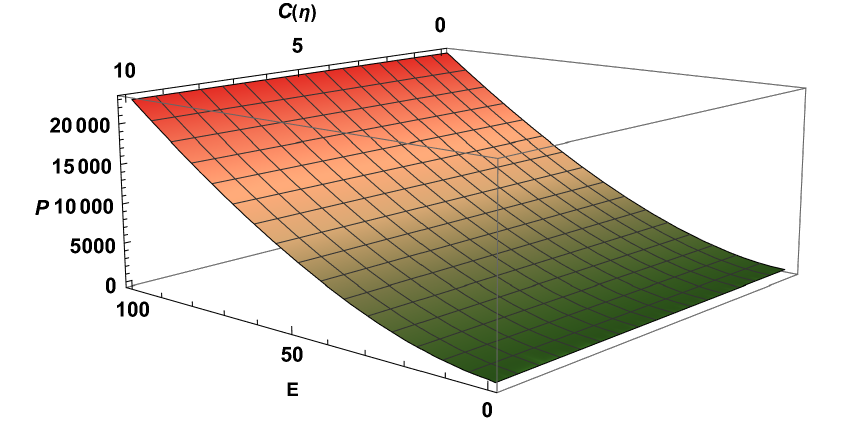}
  \includegraphics[width=8cm,height=5cm]{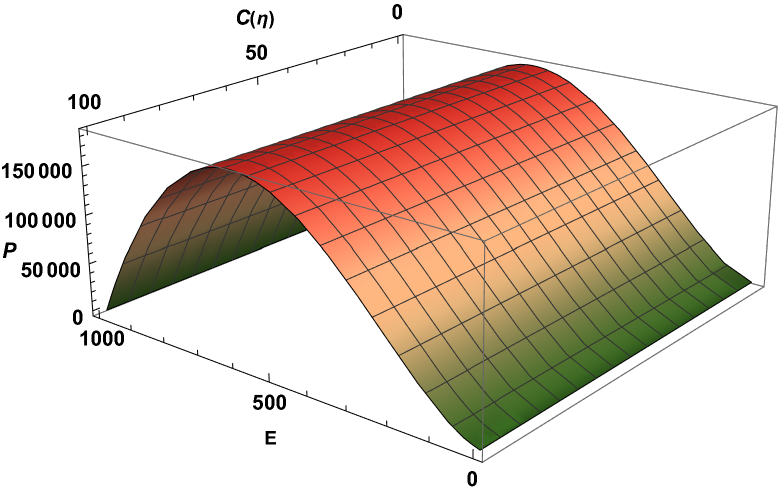}
  \includegraphics[width=8cm,height=5cm]{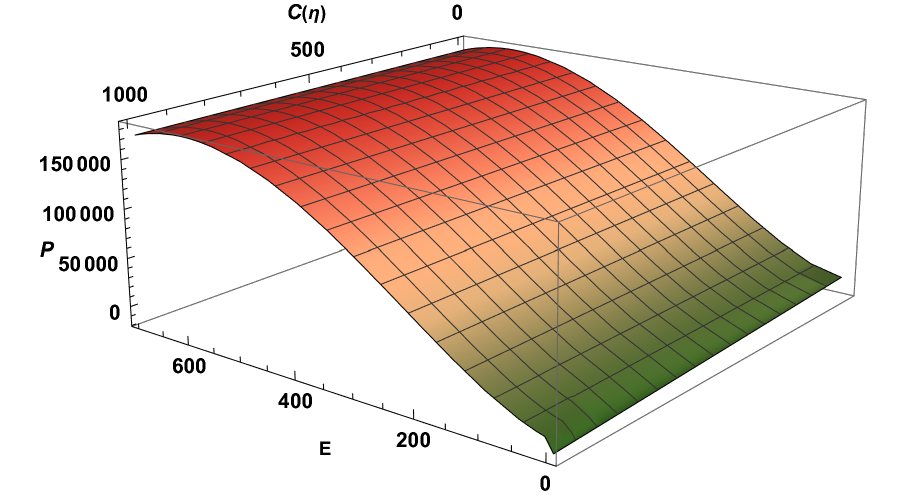}
  \caption{The modification of the equation of states due to the scale factor $C(\eta)$ considering the electroweak regime of temperature }\label{equationofstateselectroweak}
\end{figure}

\begin{figure}[tbh]
  \centering
  \includegraphics[width=8cm,height=5cm]{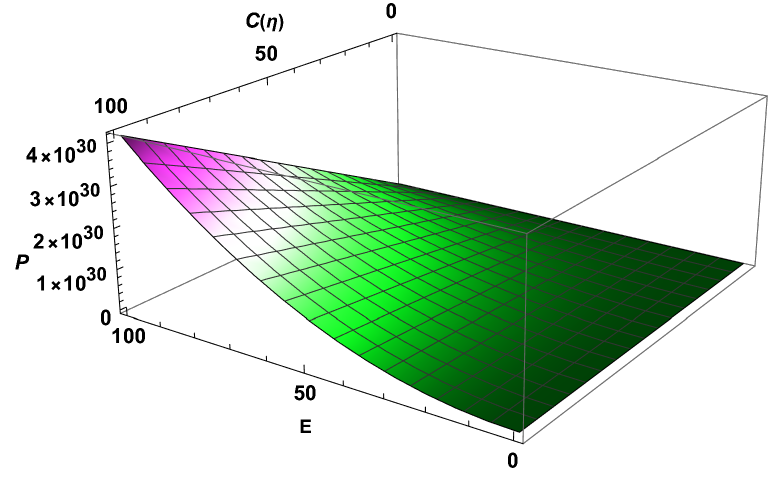}
  \includegraphics[width=8cm,height=5cm]{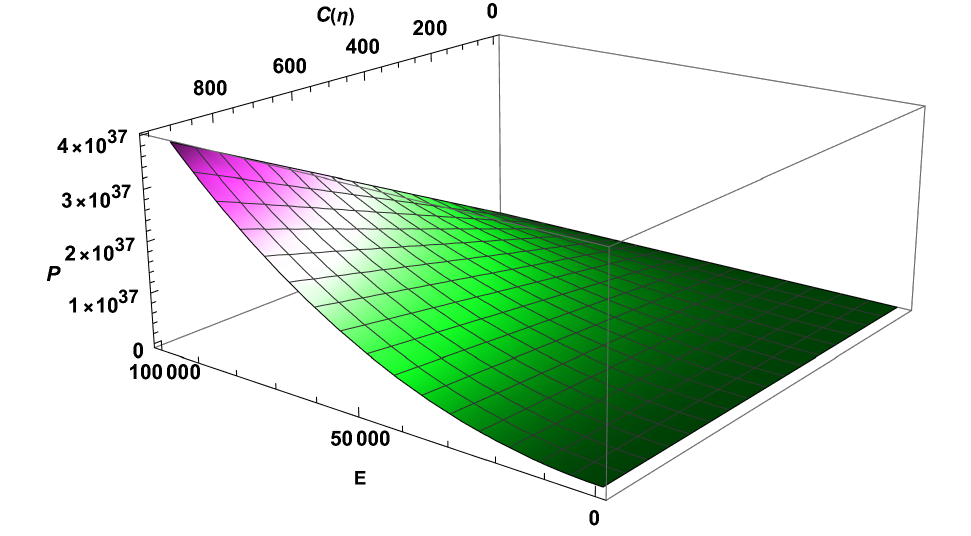}
  \includegraphics[width=8cm,height=5cm]{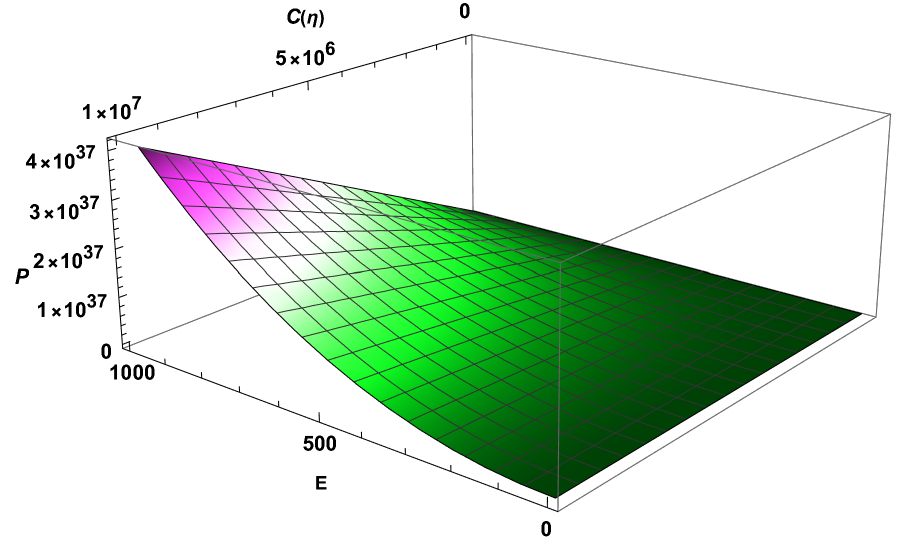}
  \caption{The modification of the equation of states due to the scale factor $C(\eta)$ considering the CMB regime of temperature}\label{equationofstatescmb}
\end{figure}

Another remarkable topic to be studied is the equation of states \cite{pathria2011statistical}; in physics, such equation of states are thermodynamic equations correlating state variables, describing the state of matter under a given set of physical conditions, e.g., volume, temperature, pressure and mean energy \cite{callen1998thermodynamics}. Furthermore, they are effective to delineate properties of gases, solids, mixed and pure substances. Also, the state of matter in the interior of stars is one well studied topic \cite{baym1971ground,chandrasekhar1994some,eliezer2001fourth,perez2015constraining,andersson2021superfluid,ozel2006soft,stergioulas2003rotating}. Initially, we consider the pressure $P$
\ie
P = \frac{1}{\beta \pi^{2}} \int^{\infty}_{0} [C(\eta) -E^{2}] \mathrm{ln}(1-e^{-\beta E}) \mathrm{d}E
\fe
which may be expanded as 
\ie
\begin{split}
P = &\frac{E^4 \left(-\beta^4 C(\eta)-120 \beta^2\right)}{2880}+\frac{1}{24} E^2 \left(\beta^2 C(\eta) -24 \mathrm{ln} (\beta  E)\right)  \\
&- \frac{1}{2} E \beta  C(\eta) + C(\eta) \mathrm{ln} (\beta  E)+\frac{\beta  E^3}{2} +\mathcal{O}\left(E^5\right).
\end{split}
\fe
Analogously, the same temperatures will be considered here in order to provide a concise analysis of the equation of states.
Considering the inflationary era, we clearly see the modification of the equation of states due to the presence of the scale factor $C(\eta)$. From Fig. \ref{equationofstatesinflation}, it indicates that the pressure tends to attenuate its values when $C(\eta)$ increases. Moreover, from Fig. \ref{equationofstateselectroweak}, we verify a huge difference between those configurations in the plots, for different values of $E, C(\eta), P$. On the other hand, when we consider the cosmic microwave background temperature displayed in Fig. \ref{equationofstatescmb}, even though possessing substantial modifications in the parameters $E, C(\eta)$, and $P$, we practically did not get much changes concerning the shape of the plots.

\subsubsection{Entropy}

\begin{figure}[tbh]
  \centering
  \includegraphics[width=8cm,height=5cm]{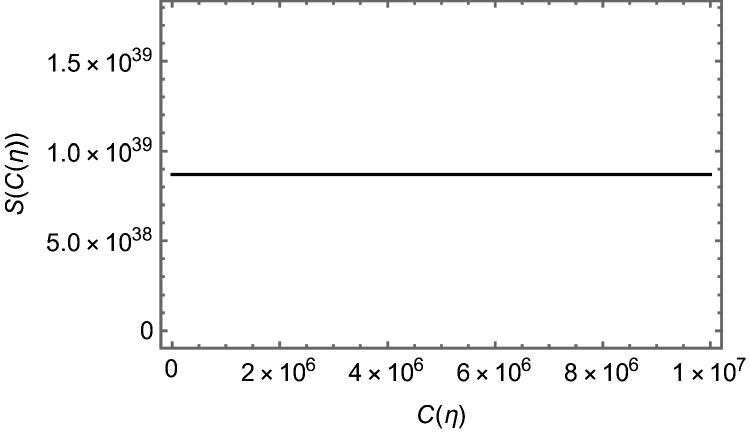}
  \includegraphics[width=8cm,height=5cm]{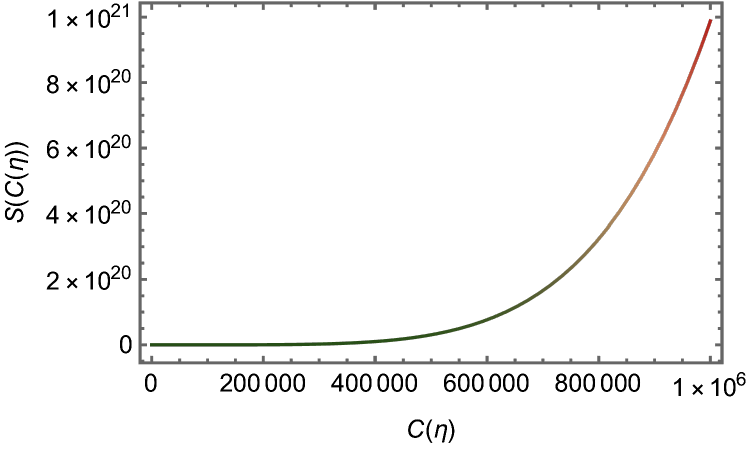}
  \includegraphics[width=8cm,height=5cm]{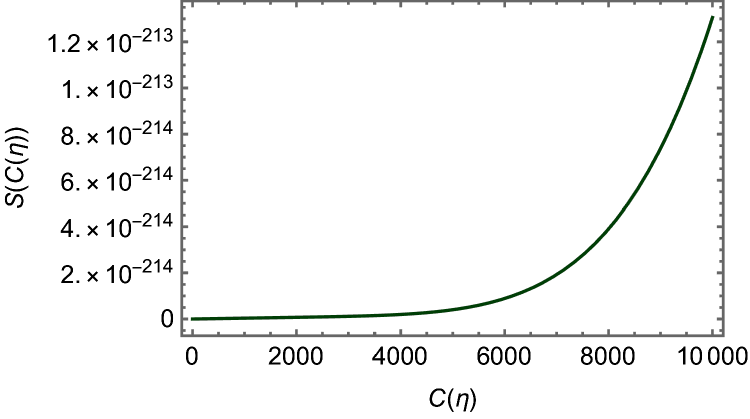}
  \caption{The entropy for different configurations of temperature: top left (inflationary era), top right (electroweak epoch), and bottom (CMB) }\label{entropies}
\end{figure}

Taking into account the first law of the thermodynamics, i.e., $Q=\Delta E + \tau$, we realize that the entropy is a fundamental feature to address many aspects of physical models. In this way, it is given by
\ie
S(\beta,C(\eta))= \int^{\infty}_{0} \frac{1}{\pi^{2}}   \left[ \beta \int^{\infty}_{0} \frac{  E \left(E^2-C(\eta)\right) e^{-\beta  E}}{ 1-e^{-\beta E}} - \int^{\infty}_{0} \left(E^2 - C(\eta)\right) \mathrm{ln} \left(1-e^{-\beta  E}\right) \right]\mathrm{d}E
\fe
The plots are shown in Fig. \ref{entropies} for different ranges of temperatures. Notice that within the inflationary approach of temperature, the entropy remains constant for different values of the scale factor $C(\eta)$. Also, for the electroweak temperature, the second law of the thermodynamics is still maintained. On the other hand, the low temperature considered here was $T=10^{-3}$ GeV instead of the cosmic microwave background; this happened because even though considering such temperature, the magnitude of the entropy was too small. If we regard a lower temperature otherwise, there would exist only the trivial contribution.

\subsubsection{Heat capacity}

\begin{figure}[tbh]
  \centering
  \includegraphics[width=8cm,height=5cm]{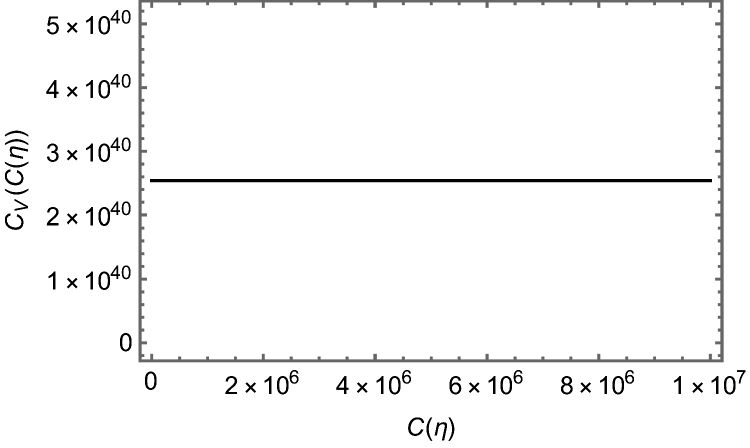}
  \includegraphics[width=8cm,height=5cm]{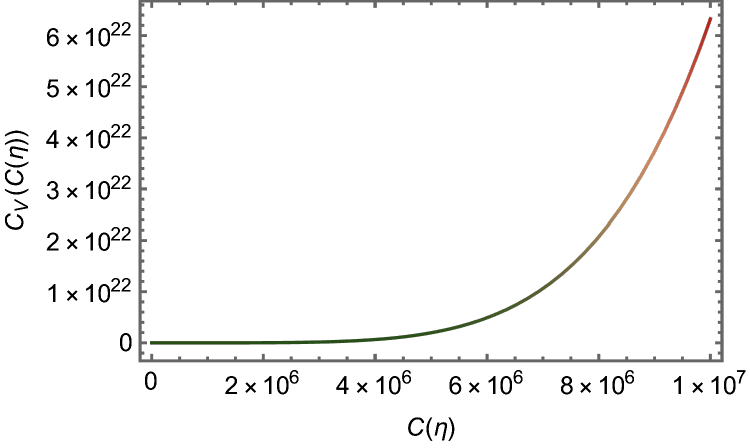}
  \includegraphics[width=8cm,height=5cm]{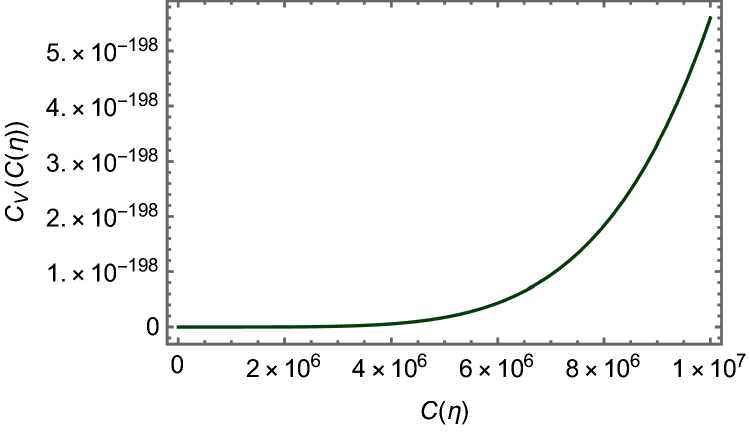}
  \caption{The heat capacity for different configurations of temperature: top left (inflationary era), top right (electroweak epoch), and bottom (CMB) }\label{heatcappacities}
\end{figure}

This section is devoted to verify the consequences of both the scale factor $C(\eta)$ and the distinct temperatures for the heat capacity. It is written as
\ie
C_{V}(\beta,C(\eta))=  \frac{ \beta^{2}}{\pi^{2}}   \left[ \int^{\infty}_{0} \frac{  E^{2} \left(E^2-C(\eta)\right) e^{-\beta  E}}{ 1-e^{-\beta E}} + \int^{\infty}_{0} E^{2}\left(E^2 - C(\eta)\right) \frac{e^{-2\beta E}}{(1-e^{-\beta E})^{2}} \right]\mathrm{d}E.
\fe
The complete analysis of the heat capacity is presented in Fig. \ref{heatcappacities}. For the inflationary era, the thermal function maintained constant for diverse values of $C(\eta)$. Furthermore, in the electroweak epoch regime of temperature, the magnitude of this thermal function increases whenever $C(\eta)$ changes. On ther other hand, it is worth mentioning that the low temperature regarded for the heat capacity was $T=10^{-3}$ GeV as well instead of the cosmic microwave background; this is because the magnitude of the heat capacity was too small in such configuration. If we had considered a lower temperature otherwise, there would exist only a trivial contribution as well.

\section{The second case: $s=1, p=1$}

As we have done in the previous section, here, we are going to perform the subsequent step toward the simplest case, i.e., we shall consider our system governed by the following configuration $s=1, p=1$. Notice that, for different values of $s, p$, we can also obtain the same dispersion relation, e.g., $s=2, p=0; s=0, p=2$. Both of them give rise to a particular set of configuration from Eq. (\ref{generaldispersionrelation}), which leads to $E^2 = {\bf{k}^2} + {\bf{k}^4}/C(\eta)$. Moreover, it is obvious to realize that this equation has four different solutions. Nevertheless, just one of them seems to fit our purpose, i.e., because it has real positive defined values. In essence, the solution is given by
\ie
{\bf{k}}_{2} = \frac{\sqrt{\sqrt{C(\eta)} \sqrt{4 E^2+C(\eta)}-C(\eta)}}{\sqrt{2}},
\fe
which possesses its differential form as
\ie
\mathrm{d} {\bf{k}}_{2}= \frac{\sqrt{2} E \sqrt{C(\eta)}}{\sqrt{4 E^2+C(\eta)} \sqrt{\sqrt{C(\eta)} \sqrt{4 E^2+C(\eta)}-C(\eta)}}.
\fe
Now, we have all elements to perform the integration over the momentum space in order to build up the accessible states of the system. After that, the obtainment of the partition function will be a straightforward task. Thereby, the accessible states of the system are written as
\ie
\Omega_{2}(C(\eta)) = \frac{1}{\pi^{2}}\int^{\infty}_{0} \frac{E \sqrt{C(\eta)} \sqrt{\sqrt{C(\eta)} \sqrt{4 E^2+C(\eta)}-C(\eta)}}{\sqrt{2} \sqrt{4 E^2+C(\eta)}} \mathrm{d} E,
\fe
that we may immediately form the partition function: 
\ie
\mathrm{ln}Z_{2}[\beta,C(\eta)] = - \frac{1}{\pi^{2}} \int^{\infty}_{0} \frac{E \sqrt{C(\eta)}\, \mathrm{ln} \left(1-e^{-\beta  E}\right) \sqrt{\sqrt{C(\eta)} \sqrt{4 E^2+C(\eta)}-C(\eta)}}{\sqrt{2} \sqrt{4 E^2+C(\eta)}} \mathrm{d} E.
\label{partitionfunction2}
\fe
The above equation is all we need to address our study of thermal description of the Einstein--aether theory in the Friedmann--Robertson--Walker spacetime. As we did in the previous section, we shall perform the analysis of the mean energy, the Helmholtz free energy, entropy, and the heat capacity. More so, the correction to the \textit{Stefan--Boltzmann} law and the equation of states are provided as well. As we shall argue, all the thermodynamic state quantities have \textit{numerical} solutions only. However, under a certain limit, our entire system turns out to behave in a such way that we can generate \textit{analytical} solutions instead. It is worth mentioning that the \textit{analytical} solutions of the thermodynamic properties of more involving systems regarding cosmological scenarios are quite rare; nevertheless, when it exists, the analysis of the results can be acquired in a more accurate way. This is what we are going to accomplish in the next sections.


\subsubsection{The spectral radiance}

\begin{figure}[tbh]
  \centering
  \includegraphics[width=8cm,height=5cm]{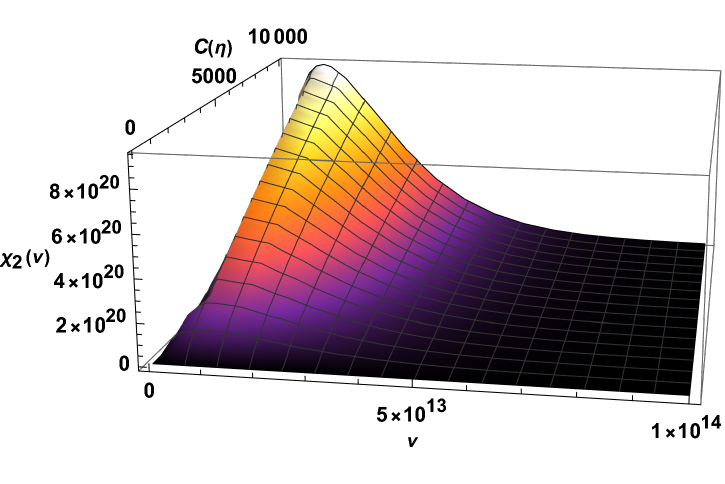}
  \includegraphics[width=8cm,height=5cm]{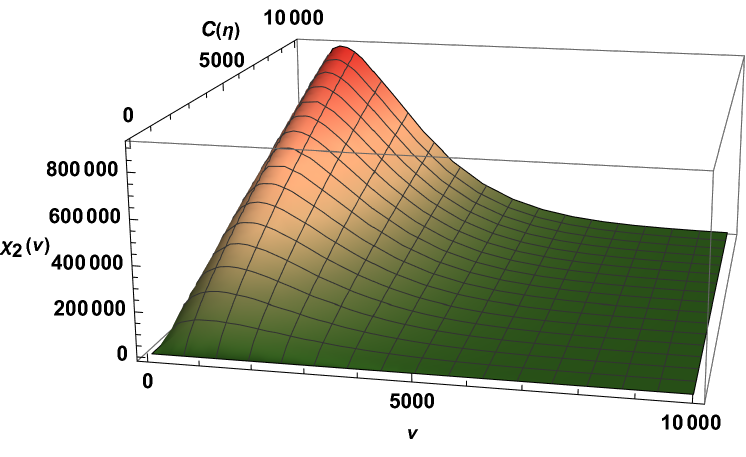}
  \caption{The figure shows the spectral radiance for different configurations of temperature. On the left, it is represented the inflationary era and, on the other hand, on the right, it is represented the electroweak epoch }\label{srmodel2}
\end{figure}

A remarkable feature whenever we are dealing with cosmological scenarios is certainly its radiation aspect. With it, we can possibly have a guide of how much of the effective power of radiation (received/emitted/transmitted) will be collected by an optical system. Such study is important since it might give rise to some indications of new phenomena which might be confronted with the observational data as soon as it is available. Specifically, we focus on photon--like particles to step forward our investigation. In this sense, from Eq. (\ref{partitionfunction2}), we can easily obtain the spectral radiance for our second model as follows
\ie
\chi_{2}(\beta,C(\eta),\nu) = \frac{(h \nu)^2 e^{-\beta  h \nu} \sqrt{C(\eta)} \sqrt{\sqrt{C(\eta)} \sqrt{4 (h \nu)^2+C(\eta)}-C(\eta)}}{\pi^{2}\sqrt{2} \left(1-e^{-\beta  h \nu}\right) \sqrt{4 (h \nu)^2+C(\eta)}}.
\label{spectralradiancemodel2}
\fe
The behavior of the spectral radiance is shown in Fig. \ref{srmodel2}, where we may verify the explicit dependence of the frequency $\nu$ and the scale factor $C(\eta)$. Particularly, we take into account two different temperatures instead, i.e., the inflationary era ($T= 10^{13}$ GeV) and the electroweak epoch ($T= 10^{3}$ GeV).
Note that the case concerning the cosmic microwave background regime of temperature ($T= 10^{-13}$ GeV) is not exhibited because the exponential factor is suppressed due to its very low values-- this gives us the trivial contribution only.

\subsubsection{Mean energy}

\begin{figure}[tbh]
  \centering
  \includegraphics[width=8cm,height=5cm]{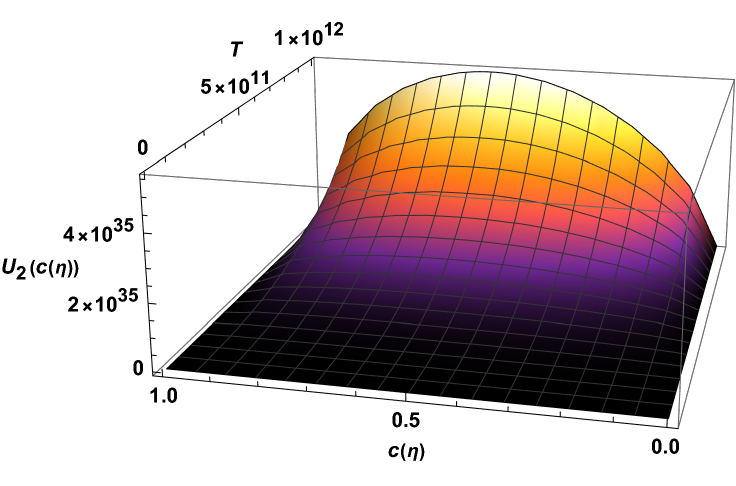}
  \includegraphics[width=8cm,height=5cm]{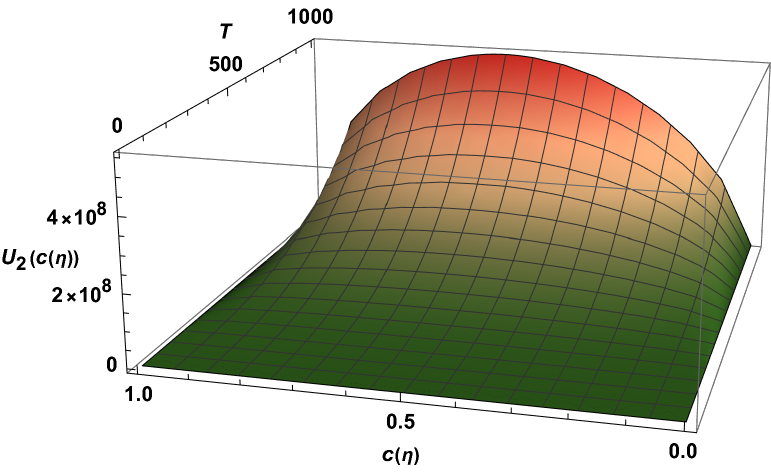}
  \includegraphics[width=8cm,height=5cm]{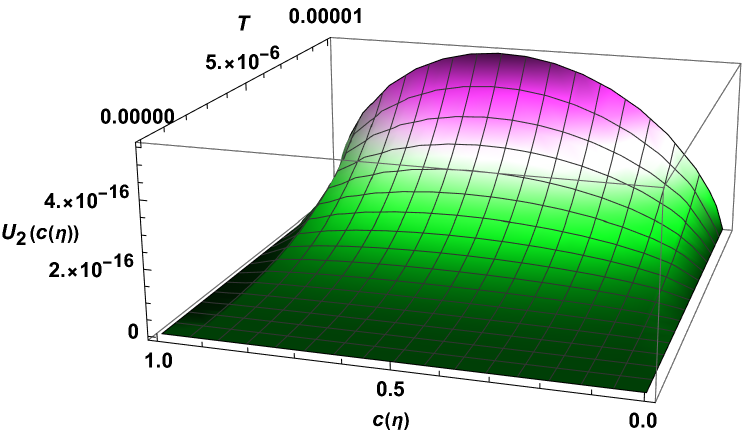}
  \caption{The mean energy for different configurations of temperature: top left (inflationary era), top right (electroweak epoch), and bottom (close to CMB) }\label{Umodel2}
\end{figure}

\begin{figure}[tbh]
  \centering
  \includegraphics[width=12cm,height=6cm]{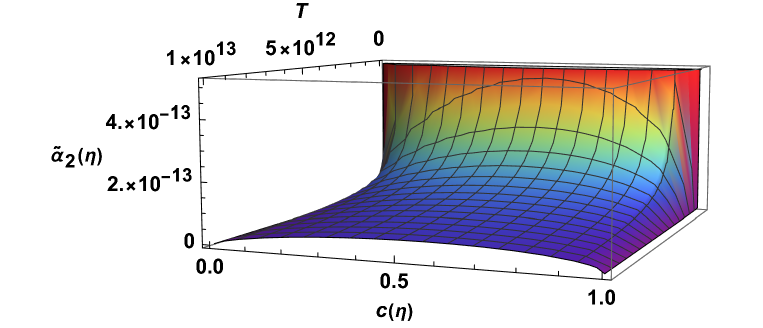}
  \caption{The correction to the \textit{Stefan--Boltzmann} law for different values of temperature }\label{sblmodel2}
\end{figure}

As it is well--known, the mean energy of a given thermal system is the energy contained within it. As a consequence, it is the necessary energy to prepare the system in any arbitrary state. In essence, it accounts for losses as well as the gains due to modifications of its internal state of energy. Furthermore, with the mean energy, we can completely describe the whole thermodynamic details of a given system. On the other hand, its value depends merely on the current state of the system; therefore, it does not depend on the particular choice of many possible processes by which energy can flow to the system. It is written as
\ie
U_{2}(\beta,C(\eta)) = \int^{\infty}_{0} \frac{E^2 e^{-\beta  E} \sqrt{C(\eta)} \sqrt{\sqrt{C(\eta)} \sqrt{4 E^2+C(\eta)}-C(\eta)}}{\pi^{2}\sqrt{2} \left(1-e^{-\beta  E}\right) \sqrt{4 E^2+C(\eta)}} \mathrm{d}E.
\label{UUmodel2}
\fe
Although it is so rare to obtain \textit{analytical} solutions for a given thermodynamic system involving cosmological scenarios, if we consider the limit where $\sqrt{4E^{2} + C(\eta)} \ll 1$, such study can properly be provided. Using this limit is entirely justified, as the parameters that govern the breaking of Lorentz symmetry are expected to be exceedingly small \cite{kostelecky2011data}. Taking into account such limit, we obtain the mean energy as 
\ie
U_{2}(\beta,C(\eta)) = \frac{\sqrt{2} \sqrt{\sqrt{C(\eta)}-C(\eta)} \sqrt{C(\eta)}}{\pi^{2}\beta ^3} \zeta (3)
\fe
where $\zeta(s)$ is the Riemann zeta function defined by
\ie
\zeta(s) = \sum^{\infty}_{n=1} \frac{1}{n^{s}} = \frac{1}{\Gamma(s)}\int_{0}^{\infty} \frac{x^{s-1}}{e^{x}-1} \mathrm{d}x, \,\,\,\,\,\,\, \Gamma(s)= \int_{0}^{\infty} x^{s-1} e^{-x} \mathrm{d}x.
\fe
The plot for the mean energy is shown in Fig. 
\ref{Umodel2}, where we see clearly how the temperature and the scale factor $C(\eta)$ modify our thermal function $U_{2}(C(\eta))$. Also, we consider three diverse temperatures, i.e., the inflationary era, the electroweak epoch, and the cosmic microwave background.
We observe a noteworthy characteristic in our findings pertaining to the theory of gravity being examined, which remains consistent across all temperature ranges: the presence of a remarkable phase transition occurring at $C(\eta)=0.5$.
As consequent investigation in this approach, the correction to the \textit{Stefan--Boltzmann} law naturally arises. It depicts the power radiated from a certain black body as a function of its temperature. To accomplish this thermal quantity, we proceed as
\ie
\tilde{\alpha}_{2}(\beta,C(\eta)) = U_{2}(\beta,C(\eta)) \beta^{4}
\fe
which, in a straightforward manner, gives 
\ie
\tilde{\alpha}_{2}(\beta,C(\eta)) = \frac{\sqrt{2} \sqrt{\sqrt{C(\eta)}-C(\eta)} \sqrt{C(\eta)}}{\pi^{2}T} \zeta (3).
\fe
The behavior of this quantity is shown in Fig. \ref{sblmodel2}.

\subsubsection{Helmholtz free energy}

\begin{figure}[tbh]
  \centering
  \includegraphics[width=8cm,height=5cm]{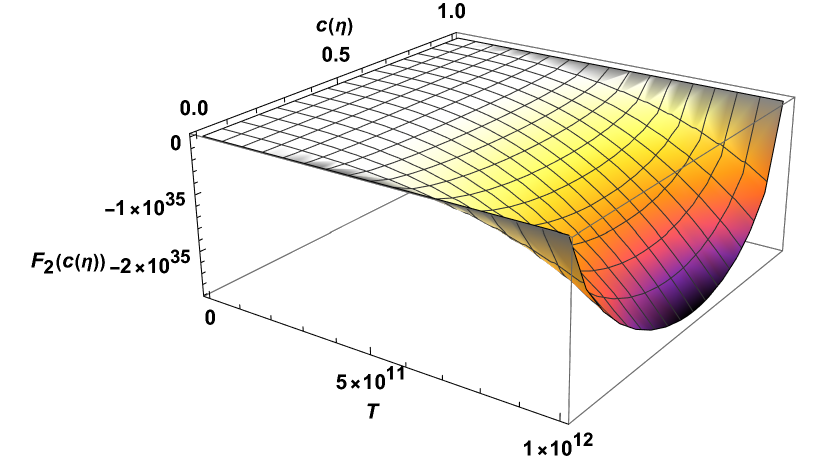}
  \includegraphics[width=8cm,height=5cm]{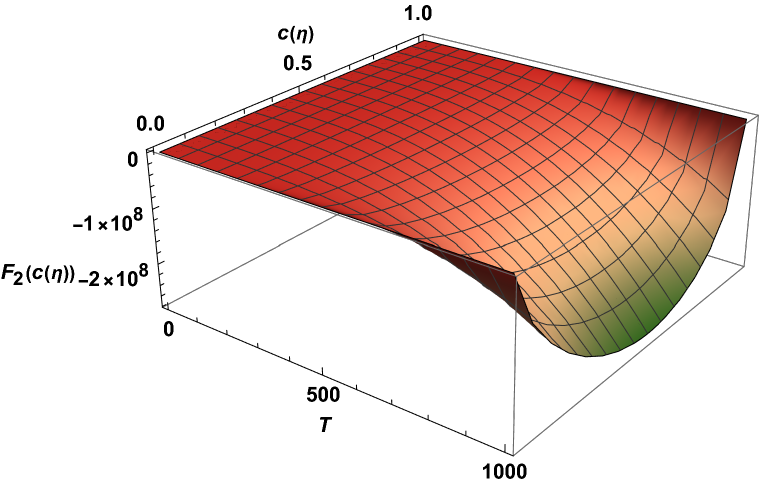}
  \includegraphics[width=8cm,height=5cm]{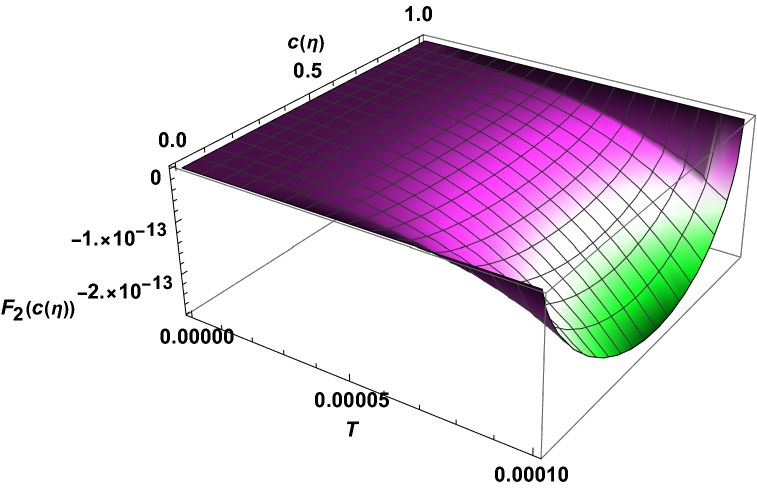}
  \caption{The Helmholtz free energy for different configurations of temperature: top left (inflationary era), top right (electroweak epoch), and bottom (close to CMB) }\label{Fmodel2}
\end{figure}

The Helmholtz free energy in thermodynamics is a thermal potential which measures the effective work gettable from a closed thermal system with constant temperature-- an isothermal process. The modification in the Helmholtz energy while a process occurs is equivalent to the maximum amount of work which a given system may carry out in a thermal process whenever the parameter temperature is maintained constant. Moreover, such thermodynamic potential is minimized at equilibrium within a constant temperature. All the features ascribed to this thermodynamic function will be provided in this section. Differently of what happened to the first case, in the second one, the following results turn out to be more robust since we obtain \textit{analytical} results. The Helmholtz free energy is given by
\ie
F_{2}(\beta,C(\eta)) = \int^{\infty}_{0} \frac{E \sqrt{C(\eta)} \mathrm{ln} \left(1-e^{-\beta  E}\right) \sqrt{\sqrt{C(\eta)} \sqrt{4 E^2+C(\eta)}-C(\eta)}}{\pi^{2}\sqrt{2} \beta  \sqrt{4 E^2+C(\eta)}} \mathrm{d} E.
\label{HHmodel2}
\fe
As one can verify, Eq. (\ref{HHmodel2}) does not have \textit{analytical} solution. Nevertheless, if we consider the limit where $\sqrt{4E^{2} + C(\eta)} \ll 1$, we are able to calculate the Helmholtz free energy analytically
\ie
F_{2}(\beta,C(\eta)) = -\frac{\zeta (3) \sqrt{\sqrt{C(\eta)}-C(\eta)} \sqrt{C(\eta)}}{\pi^{2}\sqrt{2} \beta ^3}.
\fe
For a better comprehension of the behavior of this thermodynamic state quantity, we provide a plot in which is displayed in Fig. \ref{Fmodel2}. As one can expect, there still exists the dependence of the scale factor $C(\eta)$ in our results. Furthermore, in agreement with the literature, the Helmholtz free energy decreases when the temperature increases. Particularly, we also provide the analysis taken into account those three different temperate regimes, i.e., $T=10^{13}$ GeV, $T=10^{3}$ GeV, and $T=10^{-13}$ GeV. It is worth mentioning that, in this case, the latter temperature could not be reached by the system, i.e., it was reach instead close values in comparison with the cosmic microwave background regime of temperature. In essence, with such low temperature, there was no contribution besides the trivial one.

\subsubsection{The equation of states}

\begin{figure}[tbh]
  \centering
  \includegraphics[width=8cm,height=5cm]{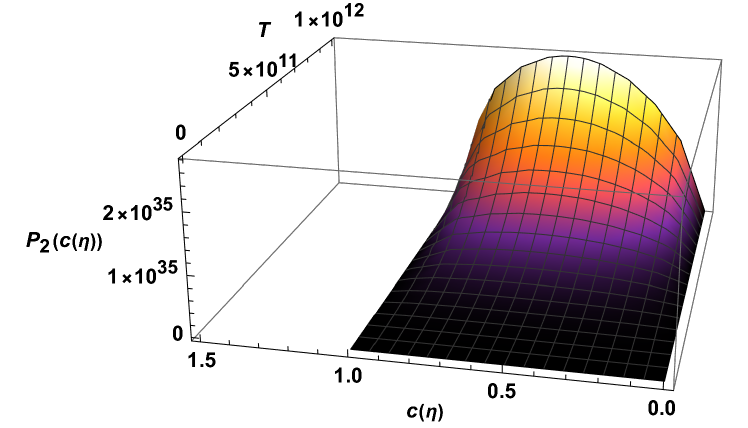}
  \includegraphics[width=8cm,height=5cm]{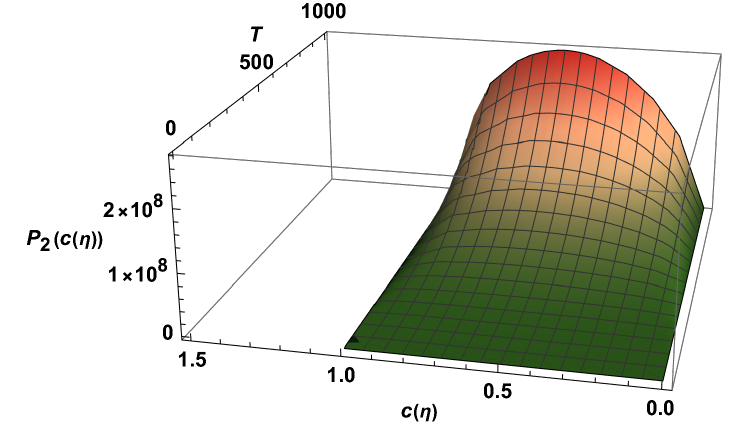}
  \includegraphics[width=8cm,height=5cm]{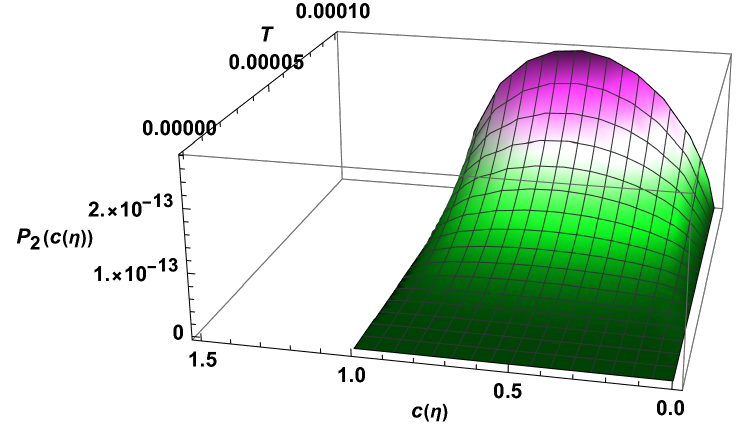}
  \caption{The equation of states for different values of temperature: top left (inflationary era), top right (electroweak epoch), and bottom (close to CMB) }\label{equationofstatesmodel2}
\end{figure}

Whenever one tries to determine fully the thermodynamic system, one inevitably deals with the equation of states of the system. In summary, it is important to have its understanding because it gives an explicitly dependence of pressure and temperature for instance. Thereby, the equation of states showing the dependence of temperature $T$, pressure $P$, and the scale factor $C(\eta)$ is written as
\ie
P_{2}(\beta,C(\eta)) = \frac{\zeta (3) \sqrt{\sqrt{C(\eta)}-C(\eta)} \sqrt{C(\eta)}}{\pi^{2}\sqrt{2}}T^{3}.
\label{pressuremodel2}
\fe
The behavior of the above equation is shown in Fig. \ref{equationofstatesmodel2} considering again three different temperatures, i.e., the inflationary era, the electroweak epoch, and the cosmic microwave background.

\subsubsection{Entropy}

\begin{figure}[tbh]
  \centering
  \includegraphics[width=8cm,height=5cm]{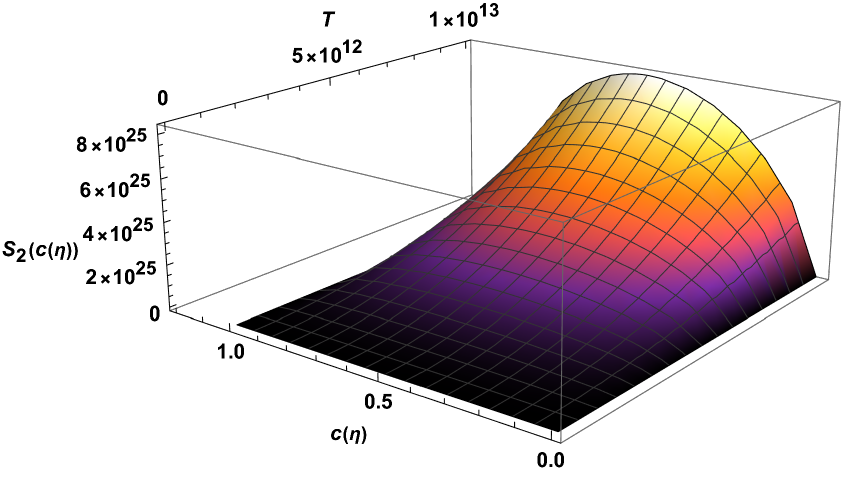}
  \includegraphics[width=8cm,height=5cm]{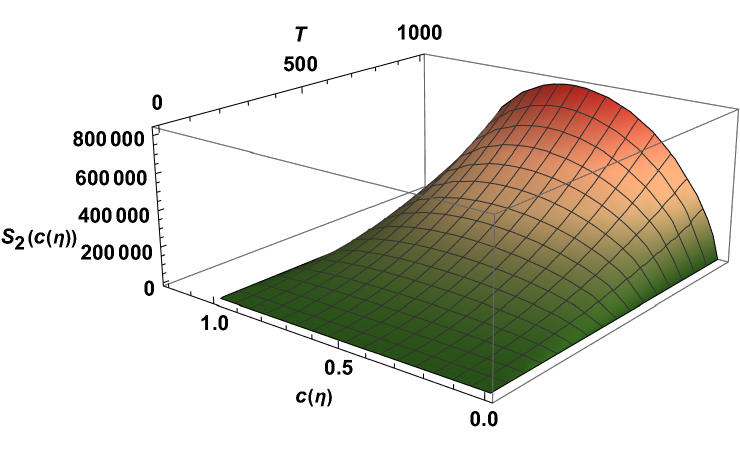}
  \includegraphics[width=8cm,height=5cm]{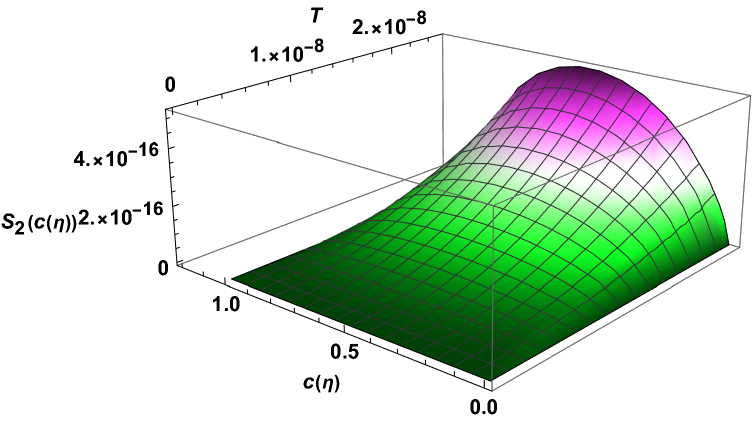}
  \caption{Entropy for different ranges of temperature: top left (inflationary era), top right (electroweak epoch), and bottom (close to CMB) }\label{SSmodel2}
\end{figure}

This is a thermodynamic quantity representing the unavailability of the energy of the a thermal system for maintaining conserved into mechanical work-- such aspect is commonly interpreted as the measure of either randomness or disorder of a given the system. In this way, the obtainment of the comprehension of such thermodynamic state quantity can give us a notable information of our system. The entropy is given by
\ie
\begin{split}
S_{2}(\beta,C(\eta)) = \frac{1}{\pi^{2}} \int^{\infty}_{0} & \beta ^2 \left(\frac{E^2 e^{-\beta  E} \sqrt{C(\eta)} \sqrt{\sqrt{C(\eta)}\sqrt{4E^{2}+C(\eta)}-C(\eta)}}{\sqrt{2} \beta  \left(1-e^{-\beta  E}\right) \sqrt{4E^{2}+C(\eta)}} \right. \\
& \left. - \frac{E \sqrt{C(\eta)} \sqrt{\sqrt{C(\eta)}\sqrt{4E^{2}+C(\eta)}-C(\eta)} \mathrm{ln} \left(1-e^{-\beta  E}\right)}{\sqrt{2} \beta ^2 \sqrt{4E^{2}+C(\eta)}}\right) \mathrm{d}E 
\end{split}.
\label{entropymodel2}
\fe
Analogously to the other thermodynamic quantities calculated so far, the entropy does not have \textit{analytical} results. Despite of this, if we consider rather the limit $\sqrt{4E^{2} + C(\eta)} \ll 1$ in Eq. (\ref{entropymodel2}), we can also acquire \textit{analytical} results and, therefore, we obtain
\ie
S_{2}(\beta,C(\eta)) = \frac{3 \zeta (3) \sqrt{\sqrt{C(\eta)}-C(\eta)} \sqrt{C(\eta)}}{\pi^{2}\sqrt{2} \beta ^2} \label{entropymodel2limit}.
\fe
The behavior of the entropy is exhibited in Fig. \ref{SSmodel2} for three different temperatures, i.e., inflationary era, electroweak epoch, and the cosmic microwave background.


\subsubsection{Heat capacity}

\begin{figure}[tbh]
  \centering
  \includegraphics[width=8cm,height=5cm]{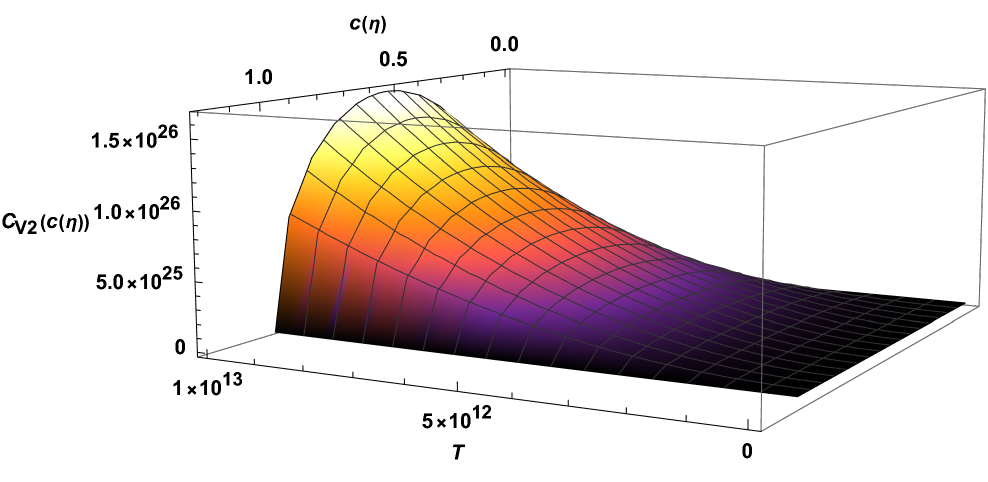}
  \includegraphics[width=8cm,height=5cm]{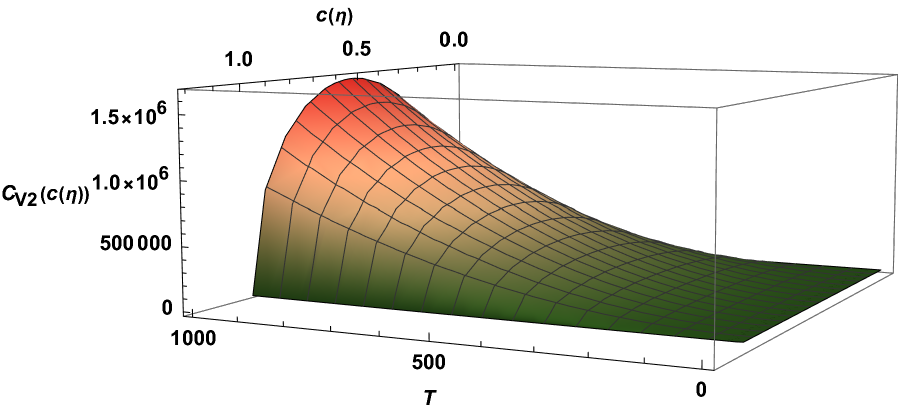}
  \includegraphics[width=8cm,height=5cm]{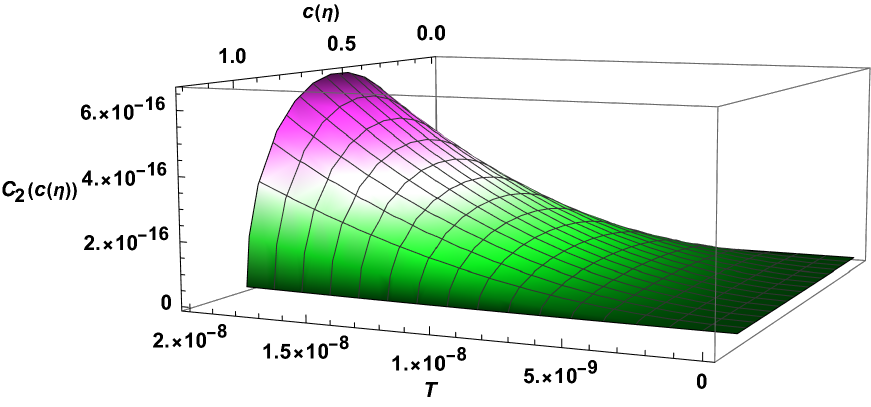}
  \caption{The Heat capacity for three configurations of temperature: top left (inflationary era), top right (electroweak epoch), and bottom (close to CMB) }\label{CCmodel2}
\end{figure}

As it is well-known the heat capacity is a physical state quantity defined as the quantity of heat to be provided to a given system to generate a unit change in its temperature. Thereby, its comprehension is quite fruitful when we are dealing with any thermal process. In this sense, the heat capacity for our second model is 
\ie
\begin{split}
C_{V2}(\beta,C(\eta)) =  \frac{1}{\pi^{2}} \int^{\infty}_{0} - & \beta ^2 \left(-\frac{E^3 e^{-\beta  E} \sqrt{C(\eta)} \sqrt{\sqrt{C(\eta)}\sqrt{4E^{2}+C(\eta)}- C(\eta)}}{\sqrt{2} \left(1-e^{-\beta  E}\right)\sqrt{4E^{2}+C(\eta)}} \right. \\
&\left.- \frac{E^3 e^{-2 \beta  E} \sqrt{C(\eta)} \sqrt{\sqrt{C(\eta)}\sqrt{4E^{2}+C(\eta)}-C(\eta)}}{\sqrt{2} \left(1-e^{-\beta  E}\right)^2 \sqrt{4E^{2}+C(\eta)}}\right) \mathrm{d}E.
\end{split}
\fe
Again, this thermodynamic quantity has only \textit{numerical} solution. However, under a certain limit (the same used in the previous section), we can solve above integral \textit{analytically}.
Then, the analytical solution to the heat capacity is
\ie
C_{V2}(\beta,C(\eta)) = \frac{3 \sqrt{2} \zeta (3) \sqrt{\sqrt{C(\eta)} - C(\eta)} \sqrt{ C(\eta)}}{\pi^{2}\beta ^2}. \label{Cmodel2}
\fe
The respective behavior of this thermal property is shown in Fig. \ref{CCmodel2} exhibiting the dependence on the temperature $T$, the scale factor $C(\eta)$ in the heat capacity. Similar to other thermodynamic quantities, the presence of a phase transition is evident at the same critical point in this context as well for all ranges of temperatures. Recently in the literature, such a feature was also expected within the context of other modified theories of gravity \cite{furtado2023thermodynamical,araujo2023thermodynamics}.

Also investigating phase transitions within the context of gravity is of utmost importance as it unveils profound insights into the intrinsic nature of spacetime and the intricate dynamics of gravitational fields. By delving into these phenomena, we enhance our comprehension of the fundamental laws that govern the universe, unraveling enigmatic phenomena like black holes and the primordial stages of the cosmos.

\section{The thrid case: $s=2, p=1$}

Without any doubt, this third case, in comparison with the first and the second ones, is more involving from a mathematical viewpoint. As we did to accomplish the analysis of the thermal aspects coming from some specific dispersion relations in the previous sections, here, we do likewise; we also proceed in order to obtain the partition function. To do it, we have to construct the accessible states of the system. Moreover, for our third study of the dispersion relation of Eq. (\ref{generaldispersionrelation}), we consider the particular case where $s=2,$ and $p=1$. As one can easily verify, such configuration generates the dispersion relation $E^2 = {\bf{k}}^2 - {\bf{k}}^6 / C(\eta)^2$, which has six distinct solutions. However, just one of them is a real positive defined:
\ie
\begin{split}
{\bf{k}}_{3}= & \left(\frac{\sqrt[3]{\frac{2}{3}} C(\eta) }{^2}{\sqrt[3]{\sqrt{3} \sqrt{27 E^4 C(\eta){}^4-4 C(\eta){}^6}-9 E^2 C(\eta){}^2}} \right.\\
&\left.  +\frac{\sqrt[3]{\sqrt{3} \sqrt{27 E^4 C(\eta){}^4-4 C(\eta ){}^6}-9 E^2 C(\eta){}^2}}{\sqrt[3]{2} 3^{2/3}}  \right)^{1/2} 
\label{k2}
\end{split}
\fe
which has the differential form as
\ie
\mathrm{d} {\bf{k}}_{3} =\frac{\frac{\frac{54 \sqrt{3} E^3 C(\eta){}^4}{\sqrt{27 E^4 C(\eta){}^4-4 C(\eta){}^6}}-18 E C(\eta){}^2}{3 \sqrt[3]{2} 3^{2/3} \left(\sqrt{3} \sqrt{27 E^4 C(\eta){}^4-4 C(\eta){}^6}-9 E^2 C(\eta){}^2\right){}^{2/3}}-\frac{\sqrt[3]{\frac{2}{3}} C(\eta){}^2 \left(\frac{54 \sqrt{3} E^3 c_{\eta }{}^4}{\sqrt{27 E^4 C(\eta){}^4-4 C(\eta){}^6}}-18 E C(\eta){}^2\right)}{3 \left(\sqrt{3} \sqrt{27 E^4 C(\eta){}^4-4 C(\eta){}^6}-9 E^2 C(\eta){}^2\right){}^{4/3}}}{2 \sqrt{\frac{\sqrt[3]{\frac{2}{3}} C(\eta){}^2}{\sqrt[3]{\sqrt{3} \sqrt{27 E^4 C(\eta){}^4-4 C(\eta){}^6}-9 E^2 C(\eta){}^2}}+\frac{\sqrt[3]{\sqrt{3} \sqrt{27 E^4 C(\eta){}^4-4 C(\eta){}^6}-9 E^2 C(\eta){}^2}}{\sqrt[3]{2} 3^{2/3}}}}.
\label{dk2}
\fe
Using Eqs. (\ref{k2}), and (\ref{dk2}) in Eq. (\ref{ms2}), we obtain the accessible states of the system
\ie
\begin{split}
\Omega_{3}(C(\eta)) =  \frac{1}{\pi^{2}}\int^{\infty}_{0} & \left(\frac{\frac{54 \sqrt{3} E^3 C(\eta){}^4}{\sqrt{27 E^4 c_{\eta }{}^4-4 C(\eta){}^6}}-18 E C(\eta){}^2}{6 \sqrt[3]{2} 3^{2/3} \left(\sqrt{3} \sqrt{27 E^4 C(\eta){}^4-4 C(\eta){}^6}-9 E^2 C(\eta){}^2\right){}^{2/3}} \right. \\
& \left.
-\frac{\sqrt[3]{\frac{2}{3}} C(\eta){}^2 \left(\frac{54 \sqrt{3} E^3 C(\eta){}^4}{\sqrt{27 E^4 C(\eta){}^4-4 C(\eta){}^6}}-18 E C(\eta){}^2\right)}{6 \left(\sqrt{3} \sqrt{27 E^4 C(\eta){}^4-4 C(\eta){}^6}-9 E^2 C(\eta){}^2\right){}^{4/3}}\right)\\
& \times
\left(\frac{\sqrt[3]{\frac{2}{3}} C(\eta){}^2}{\sqrt[3]{\sqrt{3} \sqrt{27 E^4 C(\eta){}^4-4 C(\eta){}^6}-9 E^2 C(\eta){}^2}} \right.\\
& \left. +\frac{\sqrt[3]{\sqrt{3} \sqrt{27 E^4 C(\eta){}^4-4 C(\eta){}^6}-9 E^2 C(\eta){}^2}}{\sqrt[3]{2} 3^{2/3}}\right)^{1/2} \,\mathrm{d} E.
\end{split}
\fe
In a straightforward manner, we get the partition function
\ie
\begin{split}
\mathrm{ln} Z_{3}[\beta,C(\eta)]  = - \frac{1}{\pi^{2}}\int^{\infty}_{0} & \left(\frac{\frac{54 \sqrt{3} E^3 C(\eta){}^4}{\sqrt{27 E^4 c_{\eta }{}^4-4 C(\eta){}^6}}-18 E C(\eta){}^2}{6 \sqrt[3]{2} 3^{2/3} \left(\sqrt{3} \sqrt{27 E^4 C(\eta){}^4-4 C(\eta){}^6}-9 E^2 C(\eta){}^2\right){}^{2/3}} \right. \\
& \left.
-\frac{\sqrt[3]{\frac{2}{3}} C(\eta){}^2 \left(\frac{54 \sqrt{3} E^3 C(\eta){}^4}{\sqrt{27 E^4 C(\eta){}^4-4 C(\eta){}^6}}-18 E C(\eta){}^2\right)}{6 \left(\sqrt{3} \sqrt{27 E^4 C(\eta){}^4-4 C(\eta){}^6}-9 E^2 C(\eta){}^2\right){}^{4/3}}\right) \times \mathrm{ln} (1-e^{-\beta E})\\
& \times
\left(\frac{\sqrt[3]{\frac{2}{3}} C(\eta){}^2}{\sqrt[3]{\sqrt{3} \sqrt{27 E^4 C(\eta){}^4-4 C(\eta){}^6}-9 E^2 C(\eta){}^2}} \right.\\
& \left. +\frac{\sqrt[3]{\sqrt{3} \sqrt{27 E^4 C(\eta){}^4-4 C(\eta){}^6}-9 E^2 C(\eta){}^2}}{\sqrt[3]{2} 3^{2/3}}\right)^{1/2} \,\mathrm{d} E.
\label{partitionfunction3}
\end{split}
\fe
With it, we can appropriately accomplish the investigation of thermodynamic properties for our third case analogously with we did throughout of this manuscript. In possession with the partition function, the thermal aspects turn out to be a straightforward task. In the following sections, we shall provide the thermodynamic state quantities, e.g., the mean energy, the Helmholtz free energy, pressure, entropy, and the heat capacity. The equation of states as well as the correction to the \textit{Stefan--Boltzmann} law are present as well. Differently with what happened in the previous sections, for our third case, we shall not provide the behavior of the thermodynamic function within those three different scenarios of temperature. Essentially, the reason is that for very high and very low regimes of temperatures, the integrals for these complicates expressions seem to be suppressed. Therefore, what we supply instead are some specific configurations of the system in which generate us resealable analysis for the plots concerning the thermal aspects in curved spacetime. 

\subsubsection{The spectral radiance}

\begin{figure}[tbh]
  \centering
  \includegraphics[width=8cm,height=5cm]{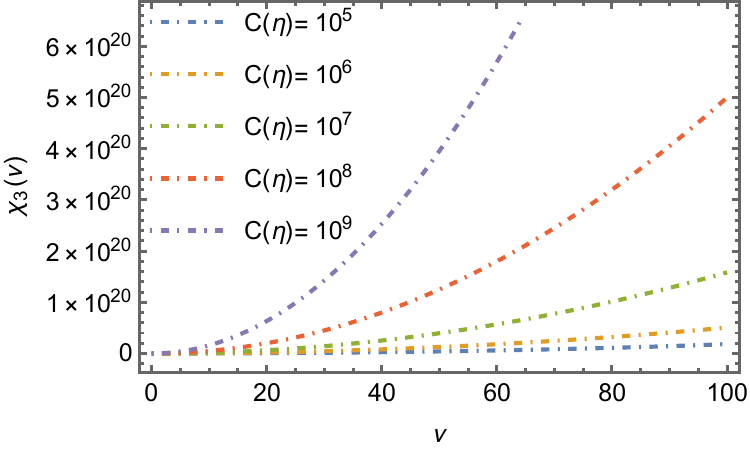}
  \includegraphics[width=8cm,height=5cm]{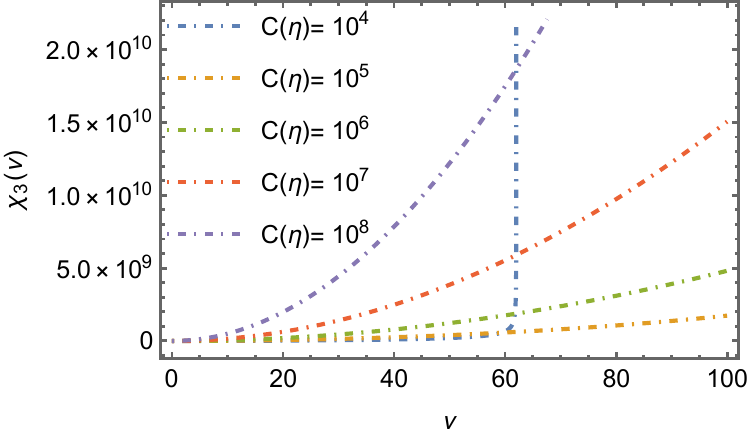}
  \includegraphics[width=8cm,height=5cm]{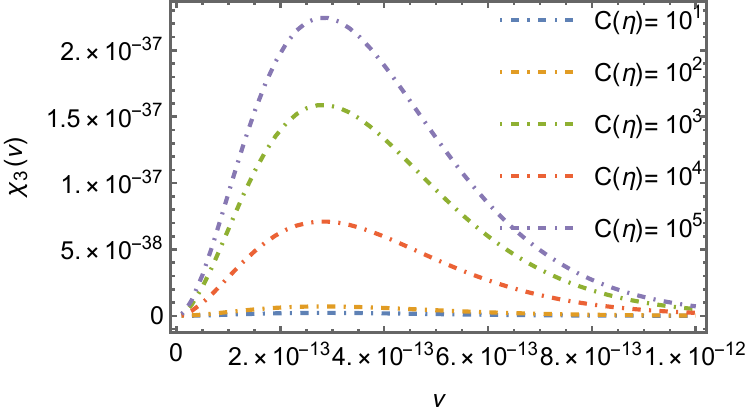}
  \caption{The spectral radiance considering different temperatures: top left (inflationary era), top right (electroweak epoch), and bottom (CMB)} \label{spectralradiancemodel3}
\end{figure}

One remarkable study that is worth taking into account is surely the spectral radiance in order to obtain more thermodynamical information about our third model. The spectral radiance can be written as
\ie
\begin{split}
\chi_{3}(\beta,C(\eta),\nu) = \frac{h \nu e^{-\beta  h \nu}}{2 \pi^{2} \left(1-e^{-\beta  h \nu}\right)}& \left(\frac{\frac{54 \sqrt{3} h \nu^3 C(\eta){}^4}{\sqrt{27 h \nu^4 C(\eta){}^4-4 C(\eta){}^6}}-18 h \nu C(\eta){}^2}{3 \sqrt[3]{2} 3^{2/3} \left(\sqrt{3} \sqrt{27 h \nu^4 C(\eta){}^4-4 C(\eta){}^6}-9 h \nu^2 C(\eta){}^2\right){}^{2/3}} \right.\\
&\left. -\frac{\sqrt[3]{\frac{2}{3}} C(\eta){}^2 \left(\frac{54 \sqrt{3} h \nu^3 C(\eta){}^4}{\sqrt{27 h \nu^4 C(\eta){}^4-4 C(\eta){}^6}}-18 h \nu C(\eta){}^2\right)}{3 \left(\sqrt{3} \sqrt{27 h \nu^4 C(\eta){}^4-4 C(\eta){}^6}-9 h \nu^2 C(\eta){}^2\right){}^{4/3}}\right) \\
& \times \left(\frac{\sqrt[3]{\frac{2}{3}} C(\eta){}^2}{\sqrt[3]{\sqrt{3} \sqrt{27 h \nu^4 C(\eta){}^4-4 C(\eta){}^6}-9 h \nu^2 C(\eta){}^2}} \right. \\
+ & \left. \frac{\sqrt[3]{\sqrt{3} \sqrt{27 h \nu^4 C(\eta){}^4-4 C(\eta){}^6}-9 h \nu^2 C(\eta){}^2}}{\sqrt[3]{2} 3^{2/3}}\right)^{1/2}.
\label{srmodel3}
\end{split}
\fe
Here, one question naturally arises: how this complicated expression in Eq. (\ref{srmodel3}) behaves for different values of temperature and frequency? This answer is provided in the plot displayed in Fig. \ref{spectralradiancemodel3}. Specially to this case, we consider three different regimes of temperature, i.e., inflationary era, electroweak epoch, and cosmic microwave background. For the first two ones, we obtain a behavior similar to the ultraviolet catastrophe-- \textit{Rayleigh-Jeans} law. On the other hand, when we take into account the cosmic microwave background regime instead, i.e., $T=10^{-13}$ GeV, a remarkable feature emerges: we obtain a genuine spectral radiation.

\subsubsection{Mean energy}

\begin{figure}[tbh]
  \centering
  \includegraphics[width=12cm,height=6cm]{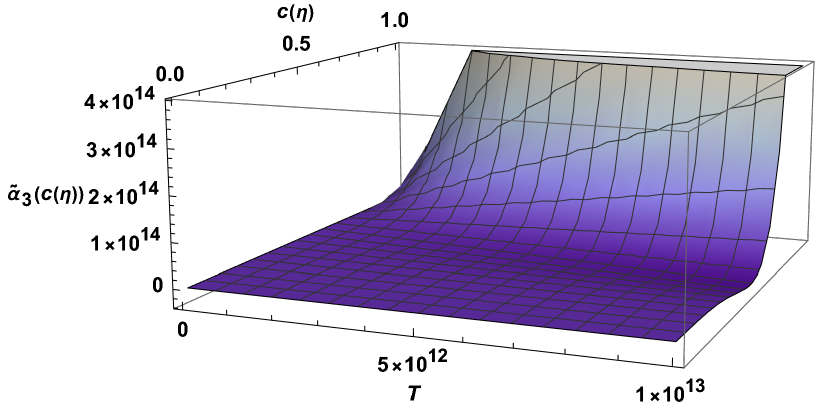}
  \caption{The correction to the \textit{Stefan--Boltzmann} law as a function of $C(\eta)$, and $T$}\label{amodel3}
\end{figure}

Initially, we start with the mean energy. Furthermore, with the partition function shown in Eq. (\ref{partitionfunction3}), we can derive it
\ie
\begin{split}
U_{3}(\beta, C(\eta)) = \frac{1}{\pi^{2}} \int_{0}^{\infty} \frac{E e^{-\beta  E}}{2 \left(1-e^{-\beta  E}\right)}& \left(\frac{\frac{54 \sqrt{3} E^3 C(\eta){}^4}{\sqrt{27 E^4 C(\eta){}^4-4 C(\eta){}^6}}-18 E C(\eta){}^2}{3 \sqrt[3]{2} 3^{2/3} \left(\sqrt{3} \sqrt{27 E^4 C(\eta){}^4-4 C(\eta){}^6}-9 E^2 C(\eta){}^2\right){}^{2/3}} \right.\\
&\left. -\frac{\sqrt[3]{\frac{2}{3}} C(\eta){}^2 \left(\frac{54 \sqrt{3} E^3 C(\eta){}^4}{\sqrt{27 E^4 C(\eta){}^4-4 C(\eta){}^6}}-18 E C(\eta){}^2\right)}{3 \left(\sqrt{3} \sqrt{27 E^4 C(\eta){}^4-4 C(\eta){}^6}-9 E^2 C(\eta){}^2\right){}^{4/3}}\right) \\
& \times \left(\frac{\sqrt[3]{\frac{2}{3}} C(\eta){}^2}{\sqrt[3]{\sqrt{3} \sqrt{27 E^4 C(\eta){}^4-4 C(\eta){}^6}-9 E^2 C(\eta){}^2}} \right. \\
+ & \left. \frac{\sqrt[3]{\sqrt{3} \sqrt{27 E^4 C(\eta){}^4-4 C(\eta){}^6}-9 E^2 C(\eta){}^2}}{\sqrt[3]{2} 3^{2/3}}\right)^{1/2} \mathrm{d}E.
\end{split}
\fe
As we can expect, above expression is much complicated having \textit{numerical} solutions only. However, under a certain limit, such thermodynamic function can be calculated \textit{analytically}. Considering the limit where $\left(\sqrt{3} \sqrt{27 E^4 C(\eta){}^4-4 C(\eta){}^6}-9 E^2 C(\eta){}^2\right)^{1/3} \ll 1$, the mean energy reads
\ie
U_{3}(\beta, C(\eta)) = \frac{1}{2 \pi^{2}} C(\eta){}^2 \left(\sqrt[3]{2}-2 \sqrt[3]{3} C(\eta){}^2\right) \sqrt{\sqrt[3]{2}+2 \sqrt[3]{3} C(\eta){}^2} \left(-\frac{2 \zeta (3)}{\beta ^3}+\frac{72 \sqrt{3} \zeta (5) C(\eta){}^2}{\beta ^5}\right).
\fe
The analysis of this thermal function is given in Fig. \ref{allmodel3}. Moreover, we can derive the correction to the \textit{Stefan--Boltzmann} law in a direct manner from the mean energy agreeing with 
\ie
\tilde{\alpha}_{3}(\beta,C(\eta)) = U(\beta,C(\eta)) \beta^{4},
\fe
which follows precisely
\ie
\tilde{\alpha}_{3}(\beta, C(\eta)) = \frac{1}{2\pi^{2}} C(\eta){}^2 \left(\sqrt[3]{2}-2 \sqrt[3]{3} C(\eta){}^2\right) \sqrt{\sqrt[3]{2}+2 \sqrt[3]{3} C(\eta){}^2} \left(-2 \zeta (3) \beta +\frac{72 \sqrt{3} \zeta (5) C(\eta){}^2}{\beta}\right).
\fe
An concise analysis taking into account different values of the scale factor $C(\eta)$, and the temperature $T$ ascribed to the correction to the \textit{Stefan--Boltzmann} law is represented in Fig. \ref{amodel3}.

\subsubsection{Helmholtz free energy}

Here, we examine how is the modification of the Helmholtz free energy considering our third model instead. As we may check, this one is a much more involving model which shows up some particularities. In this sense, the Helmholtz free energy is written as
\ie
\begin{split}
F_{3}(\beta,C(\eta)) = \int^{\infty}_{0} & \frac{1}{\pi^{2}\beta}\left(\frac{\frac{54 \sqrt{3} E^3 C(\eta){}^4}{\sqrt{27 E^4 c_{\eta }{}^4-4 C(\eta){}^6}}-18 E C(\eta){}^2}{6 \sqrt[3]{2} 3^{2/3} \left(\sqrt{3} \sqrt{27 E^4 C(\eta){}^4-4 C(\eta){}^6}-9 E^2 C(\eta){}^2\right){}^{2/3}} \right. \\
& \left.
-\frac{\sqrt[3]{\frac{2}{3}} C(\eta){}^2 \left(\frac{54 \sqrt{3} E^3 C(\eta){}^4}{\sqrt{27 E^4 C(\eta){}^4-4 C(\eta){}^6}}-18 E C(\eta){}^2\right)}{6 \left(\sqrt{3} \sqrt{27 E^4 C(\eta){}^4-4 C(\eta){}^6}-9 E^2 C(\eta){}^2\right){}^{4/3}}\right) \times \mathrm{ln} (1-e^{-\beta E})\\
& \times
\left(\frac{\sqrt[3]{\frac{2}{3}} C(\eta){}^2}{\sqrt[3]{\sqrt{3} \sqrt{27 E^4 C(\eta){}^4-4 C(\eta){}^6}-9 E^2 C(\eta){}^2}} \right.\\
& \left. +\frac{\sqrt[3]{\sqrt{3} \sqrt{27 E^4 C(\eta){}^4-4 C(\eta){}^6}-9 E^2 C(\eta){}^2}}{\sqrt[3]{2} 3^{2/3}}\right)^{1/2} \,\mathrm{d} E.
\end{split}
\fe
Now, let us take into account the same limit of other sections that gives us a precise form to the Helmholtz free energy:
\ie
F_{3}(\beta,C(\eta)) = \frac{C(\eta){}^2 \left(\sqrt[3]{2}-2 \sqrt[3]{3} C(\eta){}^2\right) \sqrt{\sqrt[3]{2}+2 \sqrt[3]{3} C(\eta){}^2} \left(\beta ^2 \zeta (3)-18 \sqrt{3} \zeta (5) C(\eta){}^2\right)}{2 \pi^{2} \beta ^5}.
\fe
For a better understanding of the detailed aspects of this thermal quantity check Fig. \ref{allmodel3}. It is worth mentioning that since we have got the Helmholtz free energy, the study of the equation of states can be derived in a direct manner.


\subsubsection{The equation of states}

One of the most important aspects that we have to take into account whenever we are dealing with a thermodynamic system is certainly the well-known equations of states. In the third model, it is evidently shown the dependence of the pressure with other thermal quantities, e.g., temperature $T$, and pressure $P$. We can write the equation of states as 
\ie
\begin{split}
P_{3}(\beta,C(\eta)) = - \int^{\infty}_{0} & \frac{1}{\pi^{2}\beta}\left(\frac{\frac{54 \sqrt{3} E^3 C(\eta){}^4}{\sqrt{27 E^4 c_{\eta }{}^4-4 C(\eta){}^6}}-18 E C(\eta){}^2}{6 \sqrt[3]{2} 3^{2/3} \left(\sqrt{3} \sqrt{27 E^4 C(\eta){}^4-4 C(\eta){}^6}-9 E^2 C(\eta){}^2\right){}^{2/3}} \right. \\
& \left.
-\frac{\sqrt[3]{\frac{2}{3}} C(\eta){}^2 \left(\frac{54 \sqrt{3} E^3 C(\eta){}^4}{\sqrt{27 E^4 C(\eta){}^4-4 C(\eta){}^6}}-18 E C(\eta){}^2\right)}{6 \left(\sqrt{3} \sqrt{27 E^4 C(\eta){}^4-4 C(\eta){}^6}-9 E^2 C(\eta){}^2\right){}^{4/3}}\right) \times \mathrm{ln} (1-e^{-\beta E})\\
& \times
\left(\frac{\sqrt[3]{\frac{2}{3}} C(\eta){}^2}{\sqrt[3]{\sqrt{3} \sqrt{27 E^4 C(\eta){}^4-4 C(\eta){}^6}-9 E^2 C(\eta){}^2}} \right.\\
& \left. +\frac{\sqrt[3]{\sqrt{3} \sqrt{27 E^4 C(\eta){}^4-4 C(\eta){}^6}-9 E^2 C(\eta){}^2}}{\sqrt[3]{2} 3^{2/3}}\right)^{1/2} \,\mathrm{d} E,
\end{split}
\fe
and under the same limit presented in the previous sections, we obtain
\ie
P_{3}(\beta,C(\eta)) = \frac{C(\eta){}^2 \left(-\sqrt[3]{2}+2 \sqrt[3]{3} C(\eta){}^2\right) \sqrt{\sqrt[3]{2}+2 \sqrt[3]{3} C(\eta){}^2} \left(-T^3 \zeta (3)+18\, T^{5}\,\sqrt{3} \zeta (5) C(\eta){}^2\right)}{2 \pi^{2}}.
\fe
There is nothing particular special when we take very high and very low regime of temperatures, e.g., either $T \rightarrow 0$ or $T \rightarrow \infty$. In summary, we have $P_{3}(\beta,C(\eta))=0$ for $T \rightarrow 0$, and $P_{3}(\beta,C(\eta))=\infty$ when $T \rightarrow \infty$ is taken into account.

On the other hand, for finishing our analysis, there exists two more thermal quantities worth to be examined: the entropy and the heat capacity. The first one can give us some important informations about the stability of set of configuration which we submit our system. On the other hand, the latter one might reveal some notable features of our thermodynamic system such as the phase transition. In order to understand these aspects of the theory, we provide the next sections as follows.

\subsubsection{Entropy}

As argued in the previous sections, there is only two more thermodynamic quantities to be investigated. Specially, we derive the entropy for our third model written as
\ie
\begin{split}
S_{3}(\beta,C(\eta)) = & \frac{1}{\pi^{2}}\int^{\infty}_{0} \frac{\beta E e^{-\beta  E}}{2   \left(1-e^{-\beta  E}\right)} \left(\frac{\frac{54 \sqrt{3} E^3 C(\eta){}^4}{\sqrt{27 E^4 C(\eta){}^4-4 C(\eta){}^6}}-18 E C(\eta){}^2}{3 \sqrt[3]{2} 3^{2/3} \left(\sqrt{3} \sqrt{27 E^4 C(\eta){}^4-4 C(\eta){}^6}-9 E^2 C(\eta){}^2\right){}^{2/3}} \right. \\
&\left. - \frac{\sqrt[3]{\frac{2}{3}} C(\eta){}^2 \left(\frac{54 \sqrt{3} E^3 C(\eta){}^4}{\sqrt{27 E^4 C(\eta){}^4-4 C(\eta){}^6}}-18 E C(\eta){}^2\right)}{3 \left(\sqrt{3} \sqrt{27 E^4v C(\eta){}^4-4 C(\eta){}^6}-9 E^2 C(\eta){}^2\right){}^{4/3}}\right) \\
&\times
\left( \frac{\sqrt[3]{\frac{2}{3}} C(\eta){}^2}{\sqrt[3]{\sqrt{3} \sqrt{27 E^4 C(\eta){}^4-4 C(\eta){}^6}-9 E^2 C(\eta){}^2}} \right. \\
&\left. + \frac{\sqrt[3]{\sqrt{3} \sqrt{27 E^4 C(\eta){}^4-4 C(\eta){}^6}-9 E^2 C(\eta){}^2}}{\sqrt[3]{2} 3^{2/3}}\right)^{1/2} \mathrm{d}E \\
& 
- \frac{1}{\pi^{2}} \int^{\infty}_{0}\frac{\mathrm{ln} \left(1-e^{-\beta  E}\right)}{2 } \left(\frac{\frac{54 \sqrt{3} E^3 C(\eta){}^4}{\sqrt{27 E^4 c_{\eta }{}^4-4 C(\eta){}^6}}-18 E C(\eta){}^2}{3 \sqrt[3]{2} 3^{2/3} \left(\sqrt{3} \sqrt{27 E^4 C(\eta){}^4-4 C(\eta){}^6}-9 E^2 C(\eta){}^2\right){}^{2/3}} \right. \\
& \left. - \frac{\sqrt[3]{\frac{2}{3}} C(\eta){}^2 \left(\frac{54 \sqrt{3} E^3 C(\eta){}^4}{\sqrt{27 E^4 C(\eta){}^4-4 C(\eta){}^6}}-18 E C(\eta){}^2\right)}{3 \left(\sqrt{3} \sqrt{27 E^4 C(\eta){}^4-4 C(\eta){}^6}-9 E^2 C(\eta){}^2\right){}^{4/3}}\right) \\
& \times \left( \frac{\sqrt[3]{\frac{2}{3}} C(\eta){}^2}{\sqrt[3]{\sqrt{3} \sqrt{27 E^4 C(\eta){}^4-4 C(\eta){}^6}-9 E^2 C(\eta){}^2}} \right. \\
& \left. + \frac{\sqrt[3]{\sqrt{3} \sqrt{27 E^4 C(\eta){}^4-4 C(\eta){}^6}-9 E^2 C(\eta){}^2}}{\sqrt[3]{2} 3^{2/3}}\right)^{1/2} \mathrm{d}E.
\end{split}
\fe
Again, using the limit $\left(\sqrt{3} \sqrt{27 E^4 C(\eta){}^4-4 C(\eta){}^6}-9 E^2 C(\eta){}^2\right)^{1/3} \ll 1$, the entropy can be \textit{analytically} derived:
\ie
S_{3}(\beta,C(\eta)) = \frac{1}{2 \pi^{2}} C(\eta){}^2 \left(\sqrt[3]{2}-2 \sqrt[3]{3} C(\eta){}^2\right) \sqrt{\sqrt[3]{2}+2 \sqrt[3]{3} C(\eta){}^2} \left(-\frac{3 \zeta (3)}{\beta ^2}+\frac{90 \sqrt{3} \zeta (5) C(\eta){}^2}{\beta ^4}\right)
\fe
The behavior of the entropy is displayed in Fig. \ref{allmodel3}.
In the next section, we devoted our attention in the remaining thermodynamic function: the heat capacity.
\subsubsection{Heat capacity}

\begin{figure}[tbh]
  \centering
  \includegraphics[width=8cm,height=5cm]{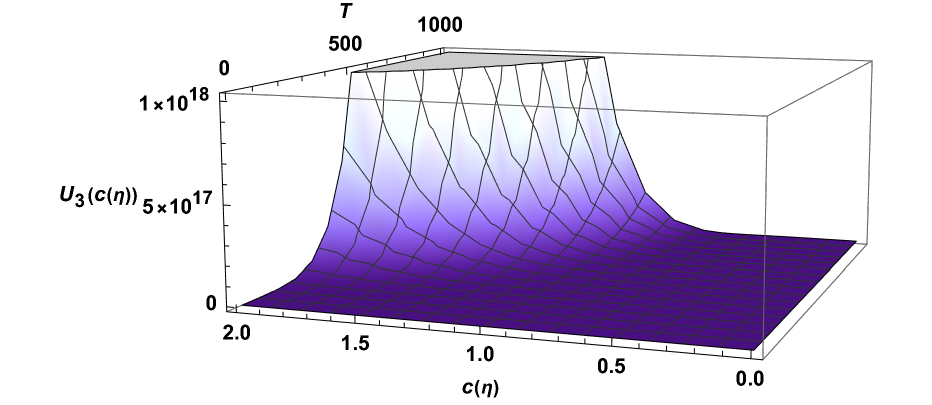}
  \includegraphics[width=8cm,height=5cm]{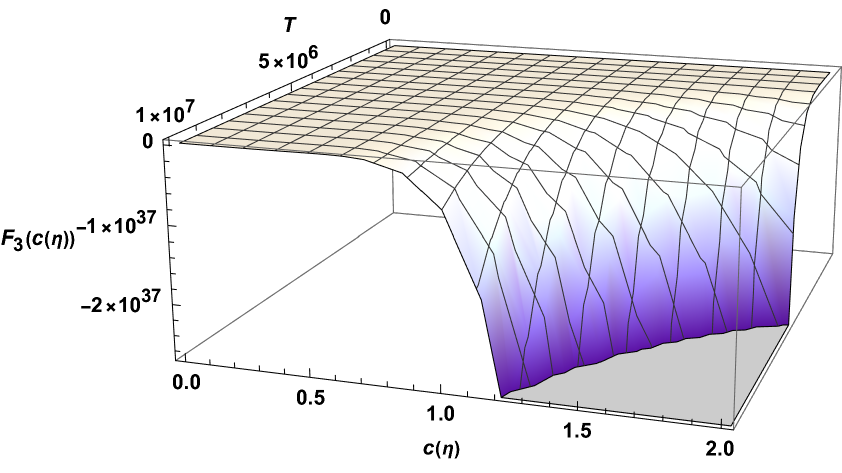}
  \includegraphics[width=8cm,height=5cm]{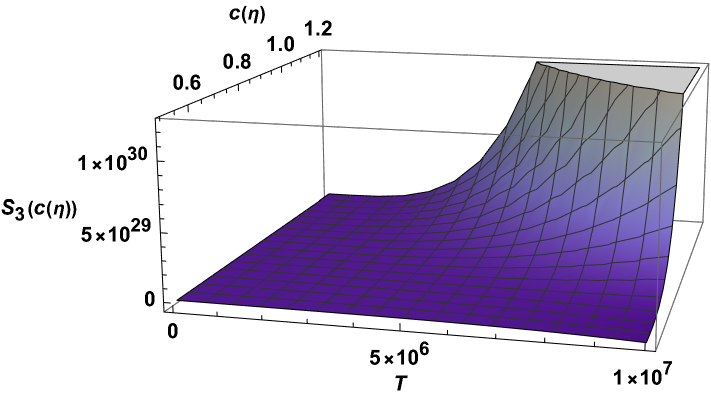}
  \includegraphics[width=8cm,height=5cm]{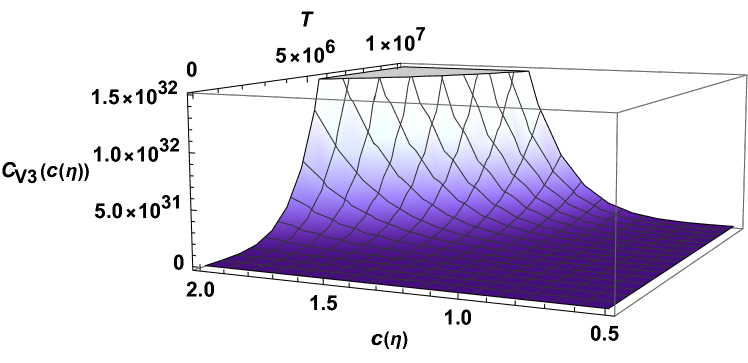}
  \caption{All thermodynamic functions are displayed: mean energy (top left), Helmholtz free energy (top right), entropy (bottom left), and heat capacity (bottom right) }\label{allmodel3}
\end{figure}

We stand with the last thermodynamic function to be analysed in this manuscript. Thereby, we can write the heat capacity as
\ie
\begin{split}
C_{V3}(\beta,C(\eta)) = &\frac{1}{\pi^{2}}\int^{\infty}_{0} \frac{\beta^{2} E^{2} e^{-2 \beta  E}}{2   \left(1-e^{-\beta  E}\right)^{2}} \left(\frac{\frac{54 \sqrt{3} E^3 C(\eta){}^4}{\sqrt{27 E^4 C(\eta){}^4-4 C(\eta){}^6}}-18 E C(\eta){}^2}{3 \sqrt[3]{2} 3^{2/3} \left(\sqrt{3} \sqrt{27 E^4 C(\eta){}^4-4 C(\eta){}^6}-9 E^2 C(\eta){}^2\right){}^{2/3}} \right. \\
&\left. - \frac{\sqrt[3]{\frac{2}{3}} C(\eta){}^2 \left(\frac{54 \sqrt{3} E^3 C(\eta){}^4}{\sqrt{27 E^4 C(\eta){}^4-4 C(\eta){}^6}}-18 E C(\eta){}^2\right)}{3 \left(\sqrt{3} \sqrt{27 E^4v C(\eta){}^4-4 C(\eta){}^6}-9 E^2 C(\eta){}^2\right){}^{4/3}}\right) \\
&\times
\left( \frac{\sqrt[3]{\frac{2}{3}} C(\eta){}^2}{\sqrt[3]{\sqrt{3} \sqrt{27 E^4 C(\eta){}^4-4 C(\eta){}^6}-9 E^2 C(\eta){}^2}} \right. \\
&\left. + \frac{\sqrt[3]{\sqrt{3} \sqrt{27 E^4 C(\eta){}^4-4 C(\eta){}^6}-9 E^2 C(\eta){}^2}}{\sqrt[3]{2} 3^{2/3}}\right)^{1/2} \mathrm{d}E \\
& 
- \frac{1}{\pi^{2}}\int^{\infty}_{0}\frac{E^{2} \beta^{2} e^{-\beta E}}{2 (1-e^{-\beta E}) } \left(\frac{\frac{54 \sqrt{3} E^3 C(\eta){}^4}{\sqrt{27 E^4 c_{\eta }{}^4-4 C(\eta){}^6}}-18 E C(\eta){}^2}{3 \sqrt[3]{2} 3^{2/3} \left(\sqrt{3} \sqrt{27 E^4 C(\eta){}^4-4 C(\eta){}^6}-9 E^2 C(\eta){}^2\right){}^{2/3}} \right. \\
& \left. - \frac{\sqrt[3]{\frac{2}{3}} C(\eta){}^2 \left(\frac{54 \sqrt{3} E^3 C(\eta){}^4}{\sqrt{27 E^4 C(\eta){}^4-4 C(\eta){}^6}}-18 E C(\eta){}^2\right)}{3 \left(\sqrt{3} \sqrt{27 E^4 C(\eta){}^4-4 C(\eta){}^6}-9 E^2 C(\eta){}^2\right){}^{4/3}}\right) \\
& \times \left( \frac{\sqrt[3]{\frac{2}{3}} C(\eta){}^2}{\sqrt[3]{\sqrt{3} \sqrt{27 E^4 C(\eta){}^4-4 C(\eta){}^6}-9 E^2 C(\eta){}^2}} \right. \\
& \left. + \frac{\sqrt[3]{\sqrt{3} \sqrt{27 E^4 C(\eta){}^4-4 C(\eta){}^6}-9 E^2 C(\eta){}^2}}{\sqrt[3]{2} 3^{2/3}}\right)^{1/2}\mathrm{d}E.
\end{split}
\fe
Notice that analogously with other ones, the heat capacity does not have \textit{analytical} solutions. Nevertheless, such issue can be overcame if the limit $\left(\sqrt{3} \sqrt{27 E^4 C(\eta){}^4-4 C(\eta){}^6}-9 E^2 C(\eta){}^2\right)^{1/3} \ll 1$ is taken into account. In this way, we have
\ie
C_{V3}(\beta,C(\eta)) = \frac{3 C(\eta){}^2 \left(\sqrt[3]{2}-2 \sqrt[3]{3} C(\eta){}^2\right) \sqrt{\sqrt[3]{2}+2 \sqrt[3]{3} C(\eta){}^2} \left(-\beta ^2 \zeta (3) + 60 \sqrt{3} \zeta (5) C(\eta){}^2\right)}{\pi^{2}\beta ^4}.
\fe
The behavior of heat capacity is shown in Fig. \ref{allmodel3}.

\section{Conclusion}

This work was aimed at providing a concise study for massless modes taking into account curved spacetime. Essentially, we explored the impact of the scale factor $C(\eta)$ coming from the Friedmann-Robertson-Walker metric within the Einstein-aether formalism to study photon-like particles. For accomplishing it, we regarded the system under the formalism of the canonical ensemble for the sake of providing the following thermodynamic state quantities: spectral radiance, Helmholtz free energy, pressure, entropy, mean energy and the heat capacity. Furthermore, the correction to the \textit{Stefan--Boltzmann} law and the equation of states were provided for all models as well. In particular, we divided our study within
three distinct cases, i.e., $s=0,p=0$; $s=1,p=1$; $s=2,p=1$. Moreover, our analyses were majority accomplished  taking into account three different regimes of temperature of the universe, i.e., the inflationary era ($T=10^{13}$ GeV), the electroweak epoch ($T=10^{3}$ GeV) and the cosmic microwave background ($T=10^{-13}$ GeV). In general, all the thermal functions had a well behavior under the variation of the temperature, i.e., the logarithm or the exponential were not suppressed in the high and low regimes of temperature.

Initially, for the first model, we considered our system being governed by the simplest case where $s=0,p=0$. All the results to this case were carried out in a \textit{numerical} approach. It is worth mentioning that for the cosmic microwave background temperature ($T=10^{-13}$ GeV), the spectral radiance seemed to show instability. For the second case, we regarded instead the configuration $s=1,p=1$. The results were derived \textit{analytically} after that the limit $\sqrt{4E^{2} + C(\eta)} \ll 1$ was taken into account. Specially, our system did not have an explicit spectral radiance to the cosmic microwave background regime of temperature. This model turned out to be in agreement with the second law of thermodynamics as one could naturally expect. On the other hand, for the third model, we considered rather $s=2,p=1$. It had more prominent challenges from the viewpoint of the power of computer calculations. More so, the spectral radiance of this one, showed a behavior similar with what happened to the \textit{Reighley--Jeans} law when we considered the inflationary era as well as the electroweak epoch temperature regimes, i.e., $T=10^{13}$ GeV and $T=10^{3}$ GeV respectively. However, when we considered instead the cosmic microwave background temperature ($T=10^{-13}$ GeV), the shape of the spectral radiance perfectly showed up. To this model, we also got \textit{analytical} solutions under a limit $\left(\sqrt{3} \sqrt{27 E^4 C(\eta){}^4-4 C(\eta){}^6}-9 E^2 C(\eta){}^2\right)^{1/3} \ll 1$. Also, all thermodynamic functions had a dependence of the Riemann zeta function $\xi(s)$. Finally, our results encountered in this manuscript did not present a dark energy--like behavior. And a phase transition was expected to our model.


\section*{Acknowledgments}
\hspace{0.5cm} This work has been supported by Conselho Nacional de Desenvolvimento Cient\'{\i}fico e Tecnol\'{o}gico (CNPq) - 142412/2018-0 and 200486/2022-5. Most of the calculations present in this manuscript were accomplished by using the \textit{Mathematica} software.

\bibliographystyle{ieeetr}
\bibliography{main}

\end{document}